\documentclass[12pt,12pt]{article}
\usepackage{graphicx}
\usepackage{epsfig}
\usepackage{amsmath,amssymb}
\usepackage{slashed} 
\catcode`\@=11
\newdimen\ex@
\font\dozeb=cmmib10 scaled \magstep1
\font\dozesyb=cmbsy10 scaled \magstep1
\font\dezb=cmmib10
\def\bm{\fam9}
\font\mathbf cmbxti10 at 12pt

\def\beq{\begin{equation}}
\def\eeq{\end{equation}}
\def\beqa{\begin{eqnarray}}
\def\eeqa{\end{eqnarray}}
\newcommand{\ba}{\begin{eqnarray}}
\newcommand{\ea}{\end{eqnarray}}
\newcommand\BA{\begin{array}}
\newcommand\EA{\end{array}}

\def\vl{{\bm  l}}

\def\vp{{\bm  p}}

\def\vv{{\bm v}}
\def\vw{{\bm w}}
\def\vx{{\bm x}}

\def\a{{\alpha }}
\def\L{{\Lambda }}
\def\r{{\rho }}
\def\A{{\bm A}}

\def\F{{\bm F}}

\def\L{{\bm L}}

\begin{document}

\title{\Large {\bf A possible quantum fluid-dynamical approach \\
\ \\[-12pt]
to vortex motion in nuclei}}
\author{Seiya NISHIYAMA\footnotemark[1]~
~and
Jo\~{a}o da PROVID\^{E}NCIA\footnotemark[1]\\
\\
Centro de F\'\i sica,
Departamento de F\'\i sica,\\
\\[-10pt]
Universidade de Coimbra,
P-3004-516 Coimbra, Portugal\footnotemark[2]\\[0.5cm]
{\it Dedicated to the Memory of Toshio Marumori}}

\def\bm#1{\mbox{\boldmath $#1$}}
\def\bra#1{\langle #1 |}
\def\ket#1{| #1 \rangle }

\maketitle

\vspace{1cm}

\footnotetext[1]{~$\!$Corresponding author. 

~~ E-mail address: 
seikoceu@khe.biglobe.ne.jp; nisiyama@teor.fis.uc.pt}
\footnotetext[2]{
$\!$  E-mail address:
providencia@teor.fis.uc.pt}

\vspace{-0.5cm}


\begin{abstract}
$\!\!\!\!\!\!\!\!\!\!$The essential point of Bohr-Mottelson theory is 
to assume a irrotational flow.
As was already suggested by Marumori
and Watanabe,
the internal rotational motion, i.e.,
the vortex motion, however, may exist
also in nuclei.
So, we must take the vortex motion
into consideration.
In classical fluid dynamics,
there are various ways to treat
the internal rotational velocity.
The Clebsch representation, 
$
\vv(\vx)
=
-\nabla \phi(\vx) + \lambda(\vx)\nabla\psi(\vx)
(\phi ; \mbox{velocity potential},
\lambda~\mbox{and}~\psi:\mbox{Clebsch parameters})
$
is very powerful
and allows for the derivation of the equations of fluid motion
from a Lagransian.
Making the best use of this advantage,
Kronig-Thellung, 
Ziman
and Ito
obtained a Hamiltonian including 
the internal rotational motion,
the vortex motion,
through the term
$\lambda(\vx) \! \nabla \! \psi(\vx)$.
Going to quantum fluid dynamics,
Ziman  and Thellung
finally derived the roton spectrum of liquid Helium II
postulated by Landau.
Is it possible to follow a similar procedure
in the description of the collective vortex motion in nuclei? 
The description of such a collective motion
has not been considered in the context of
the Bohr-Mottelson model (BMM)
for a long time.
In this paper,
we will investigate the possibility of describing
the vortex motion in nuclei
on the basis of the theories of Ziman and Ito
together with Marumori's work. 
\end{abstract}

\vspace{0.5cm}

PACS Number (s):
21.60.-n, 21.60.Ev


\newpage

\def\thesection{\arabic{section}}
\section{Introduction}

An $exact$ treatment of
collective variables in nuclei has been attempted
\cite{Maru.55,MaruYama.55,NagaTamaAmaMaru.58}.
In
\cite{Maru.55}
and
\cite{MaruYama.55},
Marumori first gave a foundation of the unified model
of collective motion and the independent particle motion in nuclei.
Applying Tomonaga's basic idea
in his collective motion theory
\cite{Tomo.55}
to nuclei, with the aid of Sunakawa's integral equation method
\cite{SYN.62},
one of the present authors (S.N.)
developed the description of nuclear surface oscillations
\cite{NishProvi.14}
and of two-dimensional nuclei
\cite{NishProvi.15}.
These descriptions are considered to provide a possible microscopic 
foundation of nuclear collective motion derived from
the Bohr-Mottelson model (BMM)
\cite{BM.74,RS.80,RoweWood.2010}.
In terms of collective variables,
these descriptions were formulated within the first 
quantized language, in contrast with the second quantized approach
in Sunakawa's method.
Extending Tomonaga's idea,
Miyazima-Tamura
\cite{MiyaTamu.56,Tamura.56}
proposed a collective 
description of nuclear surface oscillation.
An alternative attempt was proposed from a different viewpoint, 
the canonical transformation theory
\cite{Wata.56}.
To approach elementary excitations
in a one-dimensional Fermi system,
Tomonaga brought a revolutionary idea to the collective motion theory 
\cite{Tomo.55}.
The Sunakawa's method,
which is applicable also to a Fermi system,
may work well for such a problem.
It has been considered in the $exact$ canonical momenta approach
to a neutron-proton system
\cite{NishProvi2.15}.

According to Bohr-Mottelson$\!$ 
\cite{Bohr5253},
a nucleus is considered to be a portion of nuclear matter
resembling an incompressible classical fluid
but having a sharp surface.
Owing to the assumption of a small surface deformation,
collective coordinates of the surface oscillations
are described as
$
R(\theta,\varphi)
\!=\!
R_0 
\{ 
1
\!+\!
\sum_{\lambda\mu} 
\a_{\lambda\mu} Y_{\lambda\mu}(\theta,\varphi) 
\} 
(R_0:\mbox{Nuclear equilibrium radius})$.
Expanding the collective coordinates 
around equilibrium,
$\a_{\lambda\mu} \!=\! 0 $,
the surface Hamiltonian is given as
$
H 
\BA{c}
\!=\!
\sum_{\lambda\mu} \!
\left\{ \!
{\displaystyle \frac{1}{2}}B_{\lambda}|\dot{\a}_{\lambda\mu}|^2
\!+\!
{\displaystyle \frac{1}{2}}C_{\lambda}|\a_{\lambda\mu}|^2 \!
\right\} \!
\left( \!
\omega_{\lambda}
\!=\!
\sqrt{
{\frac{C_{\lambda}}{B_{\lambda}}}
} \!
\right) 
\EA
\!$
which is a set of harmonic oscillators with frequencies
$\omega_{\lambda}$.
The parameters $B_{\lambda}$ and $C_{\lambda}$ 
represent the mass parameter associated with the collective flow and
the nuclear deformability, respectively.
The essential point of their theory is 
to assume an {\bf irrotational} flow.
Let us introduce a velocity potential $\phi(\vx,t)$.
The velocity field $\vv(\vx,t) \!=\! -\nabla \phi(\vx,t)$
plays a central role over the whole theory.
The BMM, however, gives values of the moment of inertia
smaller than the empirical values
\cite{ABHMW.56}.
To resolve this,
Marumori and Watanabe
suggested to take into account
not only the surface rotation caused by the {\bf irrotational} flow
but also another kind of rotation due to the individual particle motion
\cite{Wata.56,NagaTamaAmaMaru.58}.
They called the latter rotation
the {\bf internal rotational} motion, i.e., a vortex motion.
Such a motion may be expected to occur also in nuclei.
So, it is necessary to take the vortex motion
into consideration.
In classical fluid dynamics,
there are various ways to treat
the velocity field $\vv(\vx,t)$
in connection with the {\bf internal rotational} motion.
Especially,
the Clebsch representation
\cite{Clebsch.1857}, 
$
\vv(\vx,t)
\!=\!
-\nabla \! \phi(\vx,t)
\!+\!
\lambda(\vx,t)\nabla \! \psi(\vx,t)
( \lambda, \psi\!:\!\mbox{Clebsch parameters})
$
is a useful tool
and allows for the derivation of the equations of fluid motion
from a fluid Lagransian.
Making maximum use this advantage,
Kronig-Thellung
\cite{Krothell52}, 
Ziman
\cite{Ziman53}
and Ito
\cite{Ito55}
obtained a Hamiltonian including 
the {\bf internal rotational} motion, 
the vortex motion,
via $\lambda(\vx,t)\nabla \psi(\vx,t)$
which introduces the vorticity,
$
\mbox{rot} \vv(\vx,t)
\!=\!
\nabla \lambda(\vx,t) \!\times\! \nabla \psi(\vx,t)
$.
The moment of inertia may increase by
a cooperation between {\bf rotational} and {\bf irrotational} motions.
By quantization of fluid dynamics,
Ziman and Thellung
\cite{Thell53.56}
finally derived the roton spectrum of liquid Helium II
postulated by Landau
\cite{Landau41}.
Is it possible to apply this procedure
to a description of collective vortex motion in nuclei? 
The description of such a collective motion
has not been considered in the context of the BMM
for a long time.
In this paper,
following Ziman, Ito and Marumori,
we will investigate the possibility of describing
the vortex motion in nuclei. 

In Sec. 2
we give a brief recapitulation of classical fluid dynamics in terms  of
Clebsch variables. 
In Sec. 3
we show that a unitary transformtion of an $A$-particle Hamiltonian
leads to particle and collective Hamiltonians.
The collective Hamiltonian consists of
{\bf irrotational}
and
{\bf internal rotational} motions
and their interactions. 
Section 4 is devoted to Ziman transformtion and
to derive the roton Hamiltonian.
Section 5 is devoted to determination of Clebsch parameters
through a one-form gauge potential.
Finally in Sec. 6
some discussions and further outlook are given.


\newpage

\def\thesection{\arabic{section}}
\setcounter{equation}{0}
\renewcommand{\theequation}{\arabic{section}.\arabic{equation}}     
\section{Recapitulation of classical fluid dynamics in terms  of
Clebsch representation}

For a velocity field $\vv(\vx,t)$ with a vortex component,
the Clebsch term 
$\lambda(\vx,t)\nabla\psi(\vx,t)$
is useful
to derive the equation of fluid motion
from a Hamilton formalism
\cite{Lamb,CCLin,Clebsch.1857}.
Such a representation is very effective
for passing from classical fluid dynamics to quantum fluid dynamics.
The Clebsch representation is given by\\[-10pt]
\beq
\vv(\vx,t)
=
-\nabla\phi(\vx,t)
+
\lambda(\vx,t)\nabla\psi(\vx,t) .
\label{Clebschrep}
\eeq\\[-16pt]
Then the vorticity $\vw(\vx,t)$ becomes\\[-10pt]
\beq
\vw(\vx,t)
=
\mbox{rot} \vv(\vx,t)
=
\nabla \lambda(\vx,t) \!\times\! \nabla \psi(\vx,t) ,
\label{vorticity}
\eeq\\[-16pt]
which generally does not vanish.
For simplicity,
we denote $\vv(\vx,t)$ etc. simply as $\vv$ etc.
In the Clebsch transformation
(\ref{Clebschrep}),
it is possible to choose
$\lambda$ and $\psi$
so that the surfaces
$\lambda \!=\! const.$ and $\psi \!=\! const.$
\cite{Lamb}
move with the fluid, i.e.,\\[-4pt]
\beq
\begin{array}{c}
{\displaystyle \frac{D\lambda}{Dt}}
\!=\!
\dot{\lambda}
\!+\!
\vv \!\cdot\! \nabla \! \lambda
\!=\!
0 ,~~
{\displaystyle \frac{D\psi}{Dt}}
\!=\!
\dot{\psi}
\!+\!
\vv \!\cdot\! \nabla \! \psi
\!=\!
0 ,
\end{array}
\label{substantialderivative}
\eeq\\[-10pt]
where
${\displaystyle \frac{D}{Dt}}$
is the substantial derivative
\cite{Ruthre,CCLin}.
Let us start from the Euler equation of fluid dynamics,\\[-6pt]
\beq
\int \!\! \r \frac{D\vv}{Dt} d\tau
\!=\!\!
\int \!\! \left( \r \F(\vx) \!-\! \nabla p \right) d\tau 
\rightarrow
\frac{D\vv}{Dt}
\!=\!
\F \!-\! \frac{1}{\r} \nabla p ,~
\left(
\F
\!\!=\!\!
-
\nabla U
~\mbox{and}~
\nabla P
\!\!=\!\!
\frac{1}{\r} \nabla p
\right) .
\label{Euler_equ}
\eeq \\[-8pt]
where $\rho$ and $p$ are the density and the pressure of the fluid
and
$\F$ is the external force.
The Lagrange differentiation (substantial derivative)
for the velocity,
${\displaystyle\frac{D\vv}{Dt}}$, 
is computed as \\[-14pt]
\beqa
\BA{l}
\dot{\vv}
\!+\!
(\vv \!\cdot\! \nabla)\vv
\!=\!
{\displaystyle \frac{1}{2}}
\nabla\ \!\! \vv^2
\!+\!
\dot{\vv}
\!-\!
\vv
\!\times\! \mbox{rot}\vv 
=
-\nabla \!
\left(
P
\!+\!
U
\right) ,~
\left( \! P \!=\! {\displaystyle \int}_{\!\!\!\!p_0}^p
{\displaystyle \frac{1}{\r(p)}} dp \! \right) ,
\EA
\label{Lagrangedifferentiation}
\eeqa\\[-10pt]
Assuming $\F \!=\! 0$,
with the aid of the vorticity relation
(\ref{vorticity})
and the the substantial derivatives for
$\lambda$
and
$\psi$,
(\ref{substantialderivative}),
the sum of the second and third terms in the L.H.S.
(\ref{Lagrangedifferentiation})
is calculated as\\[-14pt]
\beqa
\BA{l}
\dot{\vv}
\!-\!
\vv
\!\times\! \mbox{rot}\vv
\!=\!
-
\nabla \! \dot{\phi}
\!+\!
\dot{\lambda} \nabla \! \psi
\!+\!
\lambda \nabla \! \dot{\psi}
\!-\!
\vv \!\times\! (\nabla \! \lambda \!\times\! \nabla \! \psi )
\!=\!
-
\nabla \! \dot{\phi}
\!+\!
\left( \!
\dot{\lambda}
\!+\!
\vv \!\cdot\! \nabla \! \lambda \!
\right) \!\!
\nabla \! \lambda
\!+\!
\nabla \!\!
\left( \! \lambda \dot{\psi} \! \right) \\
\\[-10pt] 
=
-\nabla \!\!
\left( \!
\dot{\phi}
\!-\!
\lambda \dot{\psi} \!
 \right) ,
\EA
\label{Lagrangedifferentiation2}
\eeqa\\[-10pt]
from which and
(\ref{Euler_equ}),
if $U \!=\! 0$,
we have
$
\dot{\phi}
\!-\!
\lambda \dot{\psi}
\!=\!
\frac{1}{2}\vv^2
\!+\!
P
$.
Let us define a Lagrangian density\\[-6pt]
\beq
{\mathcal L}
\!=\!
\r \!
\left( \! 
\dot{\phi}
\!-\!
\lambda \dot{\psi}
\!-\!
\frac{1}{2}\vv^2 \! 
\right)
\!-\!
E_{pot}(\r) ,~
E_{pot}(\r)
\!\equiv\!
\r \!\!\! \int^{\r}_{\r_0} \!\! \frac{p \!-\! p_0}{\r^2}d\r .
\label{Lagrangian}
\eeq\\[-4pt]
The conjugate momentum to
$\phi$ and $\psi$
are expressed as
$
\pi_{\phi}
\!=\!
{\displaystyle \frac{\partial {\mathcal L}}{\partial \dot{\phi}}}
\!=\!
\r 
$
and
$
\pi_{\psi}
\!=\!
{\displaystyle \frac{\partial{\mathcal L}}{\partial \dot{\psi}}}
\!=\!
-\r \lambda .
$
Then the Hamiltonian density is given as\\[-16pt]
\beqa
\BA{l}
{\mathcal H}
\!=\!
\pi_{\phi}\dot{\phi}+\pi_{\psi}\dot{\psi}-{\mathcal L}
\!=\!
{\displaystyle \frac{1}{2}} \r \!
\left( \!
-\nabla\phi
\!-\!
{\displaystyle \frac{\pi_{\psi}}{\r}} \nabla\psi \!
\right)^{\!2}
\!+\!
E_{pot}(\r) .
\EA
\label{Hamiltonian}
\eeqa\\[-8pt]
Thus we have the canonical equations of motion
$
{\displaystyle
\dot{\r}
\!=\!
-\frac{\delta{\mathcal H}}{\delta\phi},~
\dot{\phi}
\!=\!
\frac{\delta {\mathcal H}}{\delta\r},~
\dot{\psi}
\!=\!
\frac{\delta{\mathcal H}}{\delta\pi_{\psi}}
}
$
and
$
{\displaystyle
\dot{\pi}_{\psi}
\!=\!
-\frac{\delta{\mathcal H}}{\delta \psi}
} .
$
From the first equation
we can derive the continuity equation of the fluid
as follows:\\[-10pt]
\beqa
\BA{c}
\dot{\r}
\!=\!\!
\sum_k \!
{\displaystyle \frac{\partial}{\partial x_k}} \!\!
\left( \!\!
{\displaystyle \frac{\partial {\mathcal H}}{\partial\partial _k\phi}} \!\!
\right) \!
\!=\!\!
\sum_k \!
{\displaystyle \frac{\partial}{\partial x_k}}
(-\r v_k)
\!=\!
-\mbox{div}(\r \vv) ,
(
x_k
\!=\!
x,y,z,~
\partial_k\phi
\!=\!\!
{\displaystyle \frac{\partial\phi}{\partial x_k}},
k
\!=\!
1,2,3
) .
\EA
\label{continuityEq}
\eeqa\\[-6pt]
The second, third and forth lead to
Eq.
(\ref{substantialderivative}).
Further the second also gives 
$
\dot{\phi}
\!-\!
\lambda \dot{\psi}
\!=\!
\frac{1}{2}\vv^2
\!+\!
P
$.
In the following Sections,
the classical scalar fields
$\phi (\vx,t),~\lambda (\vx,t)$ and $\psi (\vx,t)$
are treated as the corresponding quantal field operators
of the quantized fluid and
then the vector field $\vv (\vx,t)$ becomes
the quantal velocity operator.


\newpage

\def\thesection{\arabic{section}}
\setcounter{equation}{0}
\renewcommand{\theequation}{\arabic{section}.\arabic{equation}}     
\section{Unitary transformation of an $A$-particle Hamiltonian}

\vspace{-0.3cm}

In an $A$-particle system
with variables
($\vx_{n},~\!\vp_{n})$,
the Hamiltonian is given by\\[-18pt]
\beqa
\BA{c}
H
\!=\!
{\displaystyle \frac{1}{2}} \!
\sum^A_{n=1} \!
{\displaystyle \frac{\vp_n^2}{2m}}
\!+\!
V(\vx_1,\vx_2, \cdots,\vx_A) , ~(V \!\!: \mbox{Interaction potential}) .
\EA
\eeqa\\[-30pt]

For convenience,
from now on we omit the time argument.
$\!$So, here, we write $\!\rho$ etc. $\!$as $\!\rho(\vx)$ etc.
We introduce a unitary operator (UOp)
defined as the exponential of a symmetrized form of operators, \\[-18pt]
\beqa
\BA{l}
U
\!\!=\!\!
\exp \!\!
\left[ \!
- {\displaystyle \frac{i}{\hbar} \!\! \int} \!\!\!
\left\{ \!
\left( \r(\vx) \!\cdot\! \phi(\vx) \!+\! \mbox{sym.} \right)
\!\!-\!\!
\left( \r(\vx)\lambda(\vx) \!\cdot\! \psi(\vx) \!+\! \mbox{sym.} \right) \!
\right\} \! d\vx \!
\right] \! ,
\r(\vx)
\!\!=\!\!
m \!\!
\sum^A_{n=1} \!
\delta(\vx\!-\!\vx_n) ,
\EA
\eeqa\\[-14pt]
which is an extension of UOp in Ref.$\!$
\cite{MaruYama.55}
to the UOp with the Clebsch variables.
By the symbol sym. we mean that
a term of the form $Y \!\cdot\! X$ is to be added
to the operator product $X \!\cdot\! Y$.
The canonically conjugate variables $\r(\vx)$ and $\phi(\vx)$
are the density and the velocity potentials of the quantized fluid
which obey the canonical equations of motion
of the quantized fluid and satisfy the commutation relation
$[\r(\vx), \phi(\vx')] \!=\! i\hbar \delta(\vx \!-\! \vx')$.
The operators $\r(\vx)\lambda(\vx)$ and $\psi(\vx)$
satisfy a commutation relation of the same form as the above one.
As the operator $\vp_n$ is given by
$
\vp_n
\!\!=\!\!
{\frac{\hbar}{i}\frac{\partial}{\partial\vx_n}}
$, 
we have the unitary transformations of $\vp_n$
and
${ \frac{\vp_n^2}{2m}}$
as\\[-12pt]
\beqa
~~
\left\{ \!\!
\BA{l}
U\vp_n U^{-1}
=
\vp_n
\!+\!
\left[ U,~{\displaystyle \frac{\hbar}{i}\frac{\partial}{\partial\vx_n}} \right] \! 
U^{-1}
=
\vp_n
\!-\!
{\displaystyle \frac{\hbar}{i} \frac{\partial U}{\partial\vx_n}}
U^{-1} , \\
\\[-10pt]
U{\displaystyle \frac{\vp_n^2}{2m}}
U^{-1}
=
{\displaystyle \frac{\vp_n^2}{2m}}
+
\left[ U,~{\displaystyle \frac{1}{2m}}\vp_n^2 \right] \! 
U^{-1}
=
{\displaystyle \frac{\vp_n^2}{2m}}
\!+\!
{\displaystyle \frac{1}{2m} \frac{\hbar}{i}} \!
\left\{ \!
{\displaystyle 
\vp_n\frac{\partial U}{\partial \vx_n}
\!+\!
\frac{\partial U}{\partial \vx_n}\vp_n
} \!
\right\} \!
U^{-1} .
\EA
\right.
\label{unitarytransformations}
\eeqa\\[-6pt]
Owing to
$
{ 
\frac{\partial}{\partial \vx_n}
\delta \! (\vx\!-\!\vx_n)
\!=\!
-\frac{\partial}{\partial\vx}
\delta \! (\vx\!-\!\vx_n)
\!=\!
\delta \! (\vx\!-\!\vx_n) \! \frac{\partial}{\partial \vx}
}
$,
we get the gradient formula for $U$ as\\[-12pt]
\beqa
\!\!\!\!\!\!\!
\BA{lll}
{\displaystyle \frac{\partial U}{\partial \vx_n}}
&\!\!\!\!\!=\!\!\!\!\!&
-{\displaystyle \frac{i}{\hbar}} \!\!
\left[ \!
{\displaystyle \frac{\partial}{\partial \vx_n} \!\!\! \int} \!\!\!
\left\{ \!\!
\left( \!
m \!\!
\sum^A_{n'\!=\!1} \!\!
\delta(\!\vx\!-\!\vx_{n'}\!) \!\cdot\! \phi(\vx)
\!\!+\!\!
\mbox{sym.} \!
\right)
\!\!-\!\!
\left( \!
m \!\!
\sum^A_{n'\!=\!1} \!\!
\delta(\!\vx\!-\!\vx_{n'}\!)\lambda(\vx) \!\cdot\! \psi(\vx) 
\!\!+\!\!
\mbox{sym.} \!
\right) \!\!
\right\} \!\!
d\vx \!
\right] \!\!
U \\
\\[-6pt]
&\!\!\!\!\!=\!\!\!\!\!&
\left[ \!
{\displaystyle -\frac{i}{\hbar} \!\! \int} \!\!\!
\left\{ \!\!
\left( \!\!
m\delta(\vx\!-\!\vx_n)
\!\cdot\! 
{\displaystyle \frac{\partial}{\partial \vx}}
\phi(\vx)
\!\!+\!\!
\mbox{sym.} \!\!
\right)
\!\!-\!\!
\left( \!\!
m
\delta(\vx\!-\!\vx_n)
\lambda(\vx)
\!\cdot\!
{\displaystyle \frac{\partial}{\partial \vx}}
\psi(\vx) \!
\!\!+\!\!
\mbox{sym.} \!\!
\right) \!\!
\right\} \!\! d\vx \!
\right] \!\!
U. 
\EA
\label{gradientformula}
\eeqa\\[-8pt]
Note that
the gradient operator acts onto
only canonical conjugate variables
$\phi$ and $\psi$
of
$\r$ and $-\r \lambda$,
respectively.
Namely,
we here adopt
a special technical operator-action rule.
Then,
the first and second Eqs. of
(\ref{unitarytransformations})
read\\[-20pt] 
\beqa
\!\!\!\!
\left.
\BA{lll}
&\!\!\!\!&\!\!\!\!\! U\vp_n U^{-1}
\!=\!
\vp_n
\!\!+\!\!\!
{\displaystyle \int} \!\!\!
\left\{ 
\left( \!
m \delta(\vx\!-\!\vx_n)
\!\cdot\!
\nabla \! \phi(\vx)
\!+\!
\mbox{sym.} \!
\right)
\!-\!
\left( \!
\lambda(\vx) m\delta(\vx\!-\!\vx_n)
\!\cdot\!
\nabla \! \psi(\vx)
\!+\!
\mbox{sym.}
\right) 
\right\} \!
d\vx , \\
\\[-10pt]
&\!\!\!\!&\!\!\!\!\! \sum^A_{n=1}U
{\displaystyle \frac{\vp_n^2}{2m}}
U^{-1}
\!=\!
\sum_{n=1}^A \!
{\displaystyle \frac{\vp_n^2}{2m}}
\!+\!
{\displaystyle \frac{1}{2}} \!
\sum_{n\!=\!1}^A \!\!
{\displaystyle \int} \!\!
\left[
\left\{
\left( \!
\vp_n
\delta (\vx\!-\!\vx_n)
\!+\!
\delta (\vx\!-\!\vx_n) \vp_n 
\right\}
\!\cdot\!
\nabla \! \phi(\vx)
\!+\!
\mbox{sym.} \!
\right)
\right. \\
\\[-14pt]
&\!\!\!\!&\!\!\!\!\!
\left.
~~~-
\left(\!
\lambda(\vx) \!
\left\{ 
\vp_n
\delta (\vx\!-\!\vx_n)
\!+\!
\delta (\vx\!-\!\vx_n) \vp_n 
\right\} 
\!\cdot\!
\nabla \! \psi(\vx)
\!+\!
\mbox{sym.} \!
\right) 
\right] 
d\vx
\!+\!
{\displaystyle \frac{1}{2} \!\! \int} \!
\nabla \! \phi(\vx)
\!\cdot\!
\r(\vx)
\nabla \! \phi(\vx)
d\vx \\
\\[-14pt]
&\!\!\!\!-&\!\!\!\!
{\displaystyle \frac{1}{2} \!\! \int} \!\!\!
\left\{
\lambda(\vx) \!
\nabla \! \psi(\vx)
\!\cdot\!
\r(\vx) \!
\nabla \! \phi(\vx)
\!\!+\!\!
\nabla \! \phi(\vx)
\!\cdot\!
\r(\vx) \lambda(\vx) \!
\nabla \! \psi(\vx) 
\!\!-\!\!
\lambda(\vx) \!
\nabla \! \psi(\vx)
\!\cdot\!
\r(\vx) \lambda(\vx) \!
\nabla \! \psi(\vx)
\right\} \!
d\vx .
\EA \!\!\!\!
\right\}
\label{unitarytransformations2}
\eeqa\\[-22pt]

Finally, it turns out that
the original Hamiltonian is transformed to
a new Hamiltonian
$
\widetilde{H}
(\!=\!UHU^{-1})
\!\!=\!\!
\widetilde{H}_0 \!+\!\! {\displaystyle \int} \!\! \widetilde{H}_{int.} d\vx \!+\! \widetilde{H}_{field}
$
where
$
\widetilde{H}_0
\!\!=\!\!
\sum_{n=1}^A \!
{\displaystyle \frac{\vp_n^2}{2m}}
\!+\!
V(\vx_1,\cdots,\vx_A)
$
and
$\widetilde{H}_{int.}$
is given by \\[-14pt]
\beqa
\!\!\!\!
\BA{l}
\widetilde{H}_{\!int.}
\!\!=\!\!
{\displaystyle \frac{1}{2}} \!\!
\sum_{\!n\!=\!1}^A \!
\left[
\left\{ \!
\vp_n \!
\delta (\vx\!\!-\!\!\vx_n)
\!\!+\!\!
\delta (\vx\!\!-\!\!\vx_n) \vp_n \!
\right\}
\!\!\cdot\!\!
\nabla \! \phi(\vx)
\!\!-\!\!
\lambda(\vx) \!\!
\left\{ \!
\vp_n \!
\delta (\vx\!\!-\!\!\vx_n)
\!\!+\!\!
\delta (\vx\!\!-\!\!\vx_n) \vp_n \!
\right\} 
\!\!\cdot\!\!
\nabla \! \psi(\vx) \!
\!+\!
\mbox{sym.}
\right] \! .
\EA
\eeqa\\[-12pt]
The Hamiltonian
$H_{field}$ is written as
$\!{\int} \! ({\mathcal H}_{phon.}
\!+\!
{\mathcal H}_{rot.}
\!+\!
{\mathcal H}_{int.}) d\vx$
in which each Hamiltonian
${\mathcal H}$
is expressed as\\[-16pt]
\beqa
\left.
\BA{ll}
&{\mathcal H}_{phon.}
\!=\!
{\displaystyle \frac{1}{2}} \nabla \! \phi(\vx)
\!\cdot\!
\r(\vx)
\nabla \! \phi(\vx) ,~
{\mathcal H}_{rot.}
\!=\!
{\displaystyle \frac{1}{2}} 
\lambda(\vx) \!
\nabla \! \psi(\vx)
\!\cdot\!
\r(\vx) \lambda(\vx) \!
\nabla \! \psi(\vx),\\
\\[-8pt]
&{\mathcal H}_{int.}
\!=\!
-{\displaystyle \frac{1}{2}} \!
\left\{
\nabla \! \psi(\vx)
\!\cdot\!
\r(\vx) \!
\nabla \! \phi(\vx)
\!\!+\!\!
\nabla \! \phi(\vx)
\!\cdot\!
\r(\vx) \lambda(\vx) \!
\nabla \! \psi(\vx) 
\right\} .
\EA
\right\}
\label{Hfield}
\eeqa\\[-8pt]
In the Hamiltonian
for the quantized fluid,
the first and the second Hamiltonians in
(\ref{Hfield})
contribute to the occurrence of the
{\it phonon} and {\it roton} spectra,
respectively.
The last one gives their interaction.
$\!\!$They $\!$are coincident with
the classical fluid Hamiltonian$\!$
(\ref{Hamiltonian})$\!$
in the classical $\!$limit.



\newpage

\def\thesection{\arabic{section}}
\setcounter{equation}{0}
\renewcommand{\theequation}{\arabic{section}.\arabic{equation}}
\section{Ziman transformation and roton Hamiltonian}

\vspace{-0.3cm}
 
 To go from classical fluid dynamics to quantum fluid dynamics,
 Ziman introduced variables
$\psi_1\!$ and $\!\psi_2$,
$
\psi
\!\!=\!\!
{\displaystyle \frac{\psi_1}{\psi_2}} ,
\pi_{\psi}
\!\!=\!\!
-\r \lambda
\!\!=\!\!
-{\displaystyle \frac{\psi_2^2}{2}} 
\!$
and field operators
$\!
\Psi
\!\!=\!\!
{\displaystyle \frac{1}{\sqrt{\!2\hbar}}}
(\!\psi_1\!\!+\!\!i\psi_2\!)
\!$
and
$\!
\Psi^*
\!\!=\!\!
{\displaystyle \frac{1}{\sqrt{\!2\hbar}}}
(\!\psi_1\!\!-\!\!i\psi_2\!)
\!$
\cite{Ziman53}.
Introducing these relations
into
(\ref{Clebschrep})
the fluid velocity $\vv$
is expressed as\\[-24pt]
\beqa
\vv
\!=\!
-\nabla \phi
\!-\!
{\displaystyle \frac{1}{2\r}} \!
\left(
\psi_1\nabla \psi_2 \!-\! \psi_2 \nabla \psi_1 
\right)
\!=\!
-\nabla \phi
\!+\!
{\displaystyle \frac{i\hbar}{2\r}} \!
\left( \Psi^*\nabla\Psi \!-\! \Psi\nabla\Psi^* \right) , 
\label{fluidvelocity}
\eeqa\\[-18pt]
Similarly,
the Hamiltonian
${\mathcal H}_{rot.}$
in
(\ref{Hfield})
is expressed as\\[-22pt]
\beqa
\!\!
{\mathcal H}_{rot.}
\!\!=\!\!
{\displaystyle \frac{\hbar^2}{8\r}} \!\!
\left\{ \!
\left( \!
\Psi^* \nabla \Psi
\!\cdot\!
\Psi \nabla \Psi^*
\!\!+\!\!
\Psi \nabla \Psi^*
\!\cdot\!
\Psi^* \nabla \Psi \!
\right)
\!\!-\!\!
\left( \!
\Psi^{*2} 
\nabla\Psi
\!\cdot\!
\nabla\Psi
\!\!+\!\!
\Psi^2 
\nabla\Psi^*
\!\cdot\!
\nabla\Psi^* \!
\right) \!
\right\}
\!\!\equiv\!\!
{\mathcal H}_{rot.\mbox{I}}
\!+\!
{\mathcal H}_{rot.\mbox{II}} .
\label{rot.H}
\eeqa\\[-20pt]
For the incompressible fluid,
due to the continuity equation of the fluid,
we have the condition\\[-22pt]
\beqa
\mbox{div} \vv
\!=\!
-
\nabla^2 \phi
\!+\!
\frac{i\hbar}{2\r_0}
\left( \Psi^* \nabla^2 \Psi \!-\! \Psi \nabla^2 \Psi^* \right)
\!=\!
0 ,
(\r_0:\mbox{equilibrium density}).
\label{condition}
\eeqa\\[-18pt]
In the BMM,
the collective flow in nuclei is assumed to be {\bf irrotaional}.
Namely, the velocity potential satisfies
$
\nabla^2 \phi
\!=\!
0 
$
which, using
(\ref{condition}),
is ensured by requiring
$
\nabla^2 \Psi
\!=\!
\nabla^2 \Psi^*
\!=\!
0 .
$
Then the $\Psi$ and $\Psi^*$
can be expanded in term of spherical harmonics
as follows:\\[-24pt]
\beqa
\begin{array}{l}
\Psi(\vx)
\!=\!
\sum_{\lambda\mu} \!
b_{\lambda\mu} \!\!
\left( \! {\displaystyle \frac{r}{R_0}} \! \right)^{\!\!\lambda} \!\!
Y_{\lambda\mu}(\theta,\varphi) ,~
\Psi^*(\vx)
\!=\!
\sum_{\lambda\mu} \!
b_{\lambda\mu}^* \!\!
\left( \! {\displaystyle \frac{r}{R_0}} \! \right)^{\!\!\lambda} \!\!
Y_{\lambda\mu}^*(\theta,\varphi) ,
\end{array}
\label{expansionPsi}
\eeqa\\[-16pt]
where
$b_{\lambda\mu}$
and
$b_{\lambda\mu}^*$
are regarded as
boson annihilation and  creation operators
satisfying
$
[b_{\lambda\mu}, b_{\lambda'\mu'}^*]
\!\!=\!\!
\delta_{\lambda\lambda'}\delta_{\mu\mu'} ,
[b_{\lambda\mu}, b_{\lambda'\mu'}]
\!\!=\!\!
[b_{\lambda\mu}^*, b_{\lambda'\mu'}^*]
\!\!=\!\!
0 .
$
Canonical commutation relations
between $\Psi$ and $\Psi^*$
are also assumed, i.e.,
$[\Psi(\vx), \Psi^*(\vx')] \!\!=\!\! \delta(\vx \!-\! \vx')$,
$[\Psi(\vx),\Psi(\vx')]=0$ and  $[\Psi^*(\vx),\Psi^*(\vx')]=0$.
The corresponding Poisson brackets
in fluid dynamics
were discussed in detail
by Zakharov-Kuznetsov
\cite{ZakKuz.97}.
The operators
$b_{\lambda\mu}$
and
$b_{\lambda\mu}^*$
are the roton operators proposed by
Landau
\cite{Landau41}.
Ziman obtainted the roton Hamiltonian
in terms of them
and derived a roton spectrum of liquid Helium.
Ito also found  the roton Hamiltonian
by a different approach.
Adopting a vector potential $\A$
satisfying the Poisson equation
$
\nabla^2 \! \A
\!\!=\!\!
- \vw
$
for the vorticity $\vw$,
$\A$ is represented as
$
\A
\!\!=\!\!
{\displaystyle
\frac{1}{4\pi} \!\!\!
\int \!\!\!
\frac{\vw(\vx')}{|\vx \!-\! \vx'|}
d\vx'
}
$$\!$
\cite{Ito55}.
There exists an \underline{invariant integral} $I\!$,
$
I
\!\!=\!\!\!
{\displaystyle
\int \!\! \vv \!\cdot\! \vw d\vx
\!\!=\!\!
\frac{1}{4\pi} \!\!\!
\int \!\!\!\! \int \!\!\!
\frac{(\vx \!-\! \vx') \!\cdot\! [\vw(\vx) \!\times\! \vw(\vx')]}{|\vx \!-\! \vx'|^3}
d\vx d\vx'
}
$,
expressed in terms of circulations
on vortex curves $C_i$ and $C_j$
with strengths $\kappa_i$ and $\kappa_j$.
The integral $I$ may be rewritten as
$
I
\!=\!
\sum_{i,j}\alpha_{i,j} \kappa_i \kappa_j
$,
$
\alpha_{i,j}
\!\equiv\!
{\displaystyle
\frac{1}{4\pi} \!\!
\oint_{C_i}\oint_{C_j} \!\!\!
\frac{(\vx \!-\! \vx') \!\cdot\! [d\vl_i \!\times\! d\vl_j]}{|\vx \!-\! \vx'|^3}
}
~(\alpha_{i,j}\!:\underline{\mbox{winding number}}) 
$
\cite{Moffatt.69,BergerField.84,PennaSpera.89}.

First
we consider the angular momentum of
an $A$-particle system.
The total angular momentum of the system
is the sum of individual particle angular momenta.
Then we have\\[-10pt]
\beq
\BA{c}
{\mathbf J}
\!\equiv\!
\sum_{n=1}^A \!
{\mathbf j}_n 
=\!
\sum_{n=1}^A
\vx_n \!\times\! \vp_n
=
\sum_{n=1}^A \!
{\displaystyle \frac{\hbar}{i}}
\vx_n \!\times\! \nabla_n
\EA
\label{angularmomentum}
\eeq\\[-16pt]
By a unitary transformation of $\!{\mathbf J}\!$,
a new total angular momentum
$
\widetilde{\mathbf J}
(
\!=\!
U{\mathbf J}U^{-1}
)
$
is changed to\\[-22pt]
\beqa
\BA{l}
\widetilde{\mathbf J}
\!=\!
\sum_{n=1}^A \!
U {\mathbf j}_n 
U^{-1} 
\!=\!
\sum_{n=1}^A \!
\left\{ \!
{\mathbf j}_n ~\!\!\! 
\!+\!
[U, {\mathbf j}_n ]U^{-1} \!
\right\}
\!=\!
\sum_{n=1}^A \!
\left\{ \!
{\mathbf j}_n   
\!-\!
{\displaystyle \frac{\hbar}{i}}
\vx_n \!\times\! (\nabla_n U)U^{-1} \!
\right\} .
\EA
\eeqa\\[-22pt]
Here, 
$\nabla_n U$
is given by
(\ref{gradientformula}).
Define $\!\L\!$ as
$
\L
\!\equiv\!
{\displaystyle \frac{\hbar}{i}}
\vx \!\times\! \nabla
$.
Thus, we get\\[-22pt]
\beqa
\BA{l}
\widetilde{\mathbf J}
\!=\!
{\mathbf J}
\!+\!\!
{\displaystyle \int} \!
{\displaystyle \frac{i}{\hbar}} \!
\left\{ 
\r(\vx)
\L
\phi(\vx)
\!-\!
\r(\vx)\lambda(\vx)
\L
\psi(\vx) 
\right\} d\vx , 
\EA
\eeqa\\[-20pt]
which consists of  the previous operator ${\mathbf J}$
and
of the contribution to the angular momentum arising from 
the {\bf irrotational} and {\bf rotational} flows.
Thus we can find the angular momentum $\!\widehat{J}\!$
due to the {\bf rotational} flow,\\[-14pt]
\beq
\widehat{J}
\!=\!
-\frac{i}{\hbar}\r\lambda \L \psi
\!=\!
\frac{i}{\hbar}\pi_{\psi} \L \psi
\!=\!
\frac{1}{2}
\left( \Psi \L \Psi^* \!-\! \Psi^*  \L \Psi \right) ,~
\widehat{J}_{k}
\equiv
\frac{1}{2}
\left( \Psi L_k \Psi^* \!-\! \Psi^* L_k \Psi \right) .
\label{rotangularmomentum}
\eeq\\[-18pt]
The spherical tensor representation of $L_k (k\!=\!\pm,0)$
is given as 
$
L_{\pm1}
\!=\!
\!\mp\!
{\displaystyle \frac{1}{\sqrt{2}}}
(L_x \!\pm\! L_y) ,~
L_0
\!=\!
L_z .
$
Using 
the Clebsch-Gordan coefficient
$\langle l_1m_1 l_2m_2 | l_3m_3 \rangle$,
$\Psi L_k \Psi^*$ and $\Psi^* L_k \Psi$ are calculated as\\[-16pt]
\begin{eqnarray}
\BA{lll}
\Psi L_k \Psi^*
&\!\!\!\!=\!\!\!&\!\!
\sum_{\lambda'\mu'} b_{\lambda'\mu'} \!
\left(\! {\displaystyle \frac{r}{R_0}} \!\right)^{\!\!\lambda'} \!\!
Y_{\lambda'\mu'}
\sum_{\lambda\mu} b^*_{\lambda\mu} \!
\left(\! {\displaystyle \frac{r}{R_0}} \!\right)^{\!\!\lambda} \!\!
L_k Y^*_{\lambda\mu} 
\\[-4pt]
&\!\!\!\!=\!\!\!&\!\!
\sum_{\lambda'\mu'} \! \sum_{\lambda\mu}
b_{\lambda'\mu'}b^*_{\lambda\mu} \!\!
\left(\! {\displaystyle \frac{r}{R_0}} \!\right)^{\!\!\lambda+\lambda'} \!\!
Y_{\lambda'\mu'}(-1)^{\mu}L_k Y_{\lambda-\mu}
\\
&\!\!\!\!=\!\!\!&\!\!
\sum_{\lambda'\mu'} \! \sum_{\lambda\mu}
b_{\lambda'\mu'}b^*_{\lambda\mu} \!\!
\left(\! {\displaystyle \frac{r}{R_0}} \!\right)^{\!\!\lambda+\lambda'} \!\!
(-1)^{\mu+k} \!
\sqrt{\lambda(\lambda\!+\!1)}
\langle \lambda k \!-\! \mu1 \!-\! k | \lambda \!-\! \mu \rangle
Y_{\lambda'\mu'}Y_{\lambda k-\mu}
\\[-4pt]
&\!\!\!\!=\!\!\!&\!\!
\sum_{\lambda'\mu'} \! \sum_{\lambda\mu}
b_{\lambda'\mu'}b^*_{\lambda\mu}(-1)^{\mu+k} \!
\left(\! {\displaystyle \frac{r}{R_0}} \!\right)^{\!\lambda+\lambda'} \!\!\!
\sqrt{\lambda(\lambda\!+\!1)}
\langle \lambda k \!-\! \mu 1\!-\! k | \lambda \!-\! \mu \rangle
\\
&\!\!\!\!\times\!\!&\!\!\!\!
\left[
\sum_{LM} \!\!
\sqrt{{\displaystyle \frac{(2\lambda\!+\!1)(2\lambda'\!+\!1)}{4\pi(2L\!+\!1)}}}
\langle \lambda k \!-\! \mu \lambda'\mu' | LM\rangle 
\langle \lambda 0 \lambda' 0 | L 0 \rangle Y_{LM}
\right]_{M=k-\mu+\mu'} ,
\EA 
\label{Lk1}
\end{eqnarray}
\vspace{-0.6cm}
\begin{eqnarray}
\BA{lll}
\Psi^* L_k\Psi
&\!\!\!\!=\!&\!\!\!\!\!
\sum_{\lambda'\mu'} \! \sum_{\lambda\mu}
b^*_{\lambda'\mu'}b_{\lambda\mu}
\left(\! {\displaystyle \frac{r}{R_0}} \!\right)^{\!\lambda+\lambda'} \!\!\!
(-1)^{k+\mu'} \!\!
\sqrt{\lambda(\lambda\!+\!1)}
\langle \lambda \mu \!+\! k 1\!-\! k | \lambda \mu \rangle 
\\
&\!\!\!\times\!&\!\!\!\!\!
\left[
\sum_{LM} \!\!
\sqrt{{\displaystyle \frac{(2\lambda\!+\!1)(2\lambda'\!+\!1)}{4\pi(2L\!+\!1)}}}
\langle \lambda\mu \!+\! k \lambda' \!-\! \mu' | L M \rangle
\langle \lambda 0 \lambda' 0 | L 0 \rangle Y_{LM}
\right]_{M=k-\mu'+\mu} ,
\EA
\label{Lk2}
\end{eqnarray}\\[-10pt]
where
the symmetry
$
\langle \lambda \mu \!\!+\!\! k 1\!\!-\!\! k | \lambda \mu \rangle
\!=\!
(-1)^k
\langle \lambda\mu 1 k | \lambda\mu \!\!+\!\! k \rangle
$
and the formula for
the product of two spherical harmonics
are used 
\cite{Rose}.
Further using
the property of the Racah coefficients
\cite{Rose}, \\[-16pt]
\begin{eqnarray*}
\BA{cc}
&\langle \lambda\mu 1 k | \lambda\mu \!+\! k \rangle
\langle \lambda\mu \!+\! k \lambda'  \!-\!  \mu | L M \rangle \\
\\[-4pt]
&=
\sum_{L'M'} \!\!
\sqrt{(2\lambda\!+\!1)(2L'\!+\!1)}W(\lambda1 L \lambda';\lambda L')
\langle 1 k \lambda' \!-\! \mu' | L' M' \rangle
\langle \lambda\mu L' M' | L M \rangle ,
\EA
\end{eqnarray*}\\[-8pt]
and
substituting
(\ref{Lk1})
and
(\ref{Lk2})
into
(\ref{rotangularmomentum}),
the $k$-th component of {\bf rotational} angular momentum
$\widehat{\mathbf J}_{k}$
is derived as\\[-22pt]
\begin{eqnarray}
\BA{lll}
\widehat{\mathbf J}_{k}
&\!\!\!=\!\!&\!\!
{\displaystyle \frac{1}{2}} \!
\sum_{\lambda\mu\lambda'\mu'} \! \sum_{LML'M'}
(-1)^{\!\lambda+\lambda'+1}
(2\lambda\!+\!1)
\sqrt{{\displaystyle \frac{\lambda(\lambda\!+\!1)(2\lambda'\!+\!1)(2L'\!+\!1)}{12\pi}}} \!
\left( \! {\displaystyle \frac{r}{R_0}} \! \right)^{\!\lambda+\lambda'} \\
\\[-8pt]
&&\times
W( \lambda 1 L L';\lambda \lambda')
\langle \lambda 0 \lambda' 0 | L 0 \rangle Y_{LM} 
\langle L M \lambda\mu | L' M' \rangle \!
\langle L' M' \lambda' \mu' |1 k \rangle \\
\\[-8pt]
&&\times\!
\left[
b^*_{\lambda \mu}(-1)^{\mu'} \! b_{\lambda'\!-\!\mu'}
\!-\!
b^*_{\lambda' \mu'}(-1)^{\mu} b_{\lambda\!-\!\mu}
\!+\!
\delta_{\lambda \lambda'}(-1)^{\mu'} \! \delta_{\mu\!-\!\mu'}
\right] ,
\EA
\label{angmomentumJk}
\end{eqnarray}\\[-10pt]
whose form,
neglecting the constant term,
is very similar to the angular momentum
given in terms of quadratic surface-phonon operators
in the BMM.

Finally,
using the gradient formula
\cite{Edmonds,Rose}, \\[-16pt]
\begin{eqnarray*}
\nabla_k \!
\left( \! \frac{r}{R_0} \! \right)^{\!\!\lambda} \!
Y_{\lambda\mu}
&\!\!\!\!=\!\!\!\!&
-\sqrt{\frac{\lambda}{2\lambda\!-\!1}}
\langle \lambda\mu 1 k | \lambda\!-\!1 \mu\!+\!k \rangle
\frac{2\lambda\!+\!1}{R_0} \!
\left( \! \frac{r}{R_0} \! \right)^{\!\!\lambda-1} \!\!
Y_{\lambda-1\mu+k} ,
\end{eqnarray*}\\[-10pt]
and noticing the fact that
the scalar product of any two first-rank tensors
$\vv$ and $\vv'$ is given in the spherical tensor representation as
$\sum_k (-1)^k v_{1k} v'_{1-k}$,
we obtain the roton Hamiltonian
$
{\mathcal H}_{rot.}
(
\!\equiv\!
{\mathcal H}_{rot.\mbox{I}}
\!+\!
{\mathcal H}_{rot.\mbox{II}}
)
$
(\ref{rot.H})
in terms of the roton operators 
$b_{\lambda\mu}$
and
$b_{\lambda\mu}^*$
as, \\[-14pt]
\beqa
\!\!\!\!\!\!\!\!\!\!\!\!\!\!
\BA{lll}
&&\!\!\!\!\!\!\!\!\!\!{\mathcal H}_{rot.\mbox{I}}
\!\!=\!\!
{\displaystyle \frac{\hbar^2}{8\r_0}} \!\!
\sum_k \!
\sum_{\lambda\lambda'\kappa\kappa'\mu\mu'\nu\nu'} \!
\sum_{LL'MM'\Gamma\Gamma'\Lambda\Lambda'} \!
\sum_{JK}(\!-\!1)^{k\!+\!J} \!
{\displaystyle \frac{1}{R_0^2}} \!
\left(\!\!
{\displaystyle \frac{r}{R_0}} \!\!
\right)^{\!\!\lambda\!+\!\lambda'\!+\!\kappa\!+\!\kappa'\!-\!2} \!\!
{\displaystyle \frac{(2\lambda'\!\!+\!\!1) \! (2\kappa'\!\!+\!\!1)}{3}} \\
\\[-8pt]
~~~&\times&\!\!\!\!\!\!
\sqrt{\!
{\displaystyle
\frac{(2\lambda\!\!+\!\!1)\lambda'(2\lambda'\!\!-\!\!1)(2\Gamma\!\!+\!\!1)}
{4\pi}}
}
{\displaystyle
\sqrt{\! 
\frac{(2\kappa\!\!+\!\!1)\kappa'(2\kappa'\!\!-\!\!1)
(2\Gamma'\!\!+\!\!1)}
{4\pi}
}
}
\sqrt{\!{\displaystyle
\frac{(2L\!\!+\!\!1)(2L'\!\!+\!\!1)}{4\pi(2J\!\!+\!\!1)}}} \\
\\[-8pt]
&\times&\!\!\!\!\!\!
\langle \lambda 0 \lambda'\!\!-\!\!1 0 | L 0 \rangle \!
\langle \kappa 0 \kappa'\!\!-\!\!1 0 | L' 0 \rangle \!
\langle L 0 L' 0 | J 0 \rangle \!
\langle L \!M \lambda' \mu' | \Gamma \Lambda \rangle \!
\langle L' \!M' \kappa' \nu' | \Gamma' \Lambda' \rangle \!
\langle L \!M \!L' \!M' | \!J \!K \rangle \!
Y_{\!J\!K} \\
\\[-6pt]
&\times&\!\!\!\!\!\!
\langle \Gamma \Lambda \lambda \mu |1 k \rangle 
\langle \Gamma' \Lambda' \kappa \nu |1\!\!-\!\!k \rangle 
W(\lambda' \lambda'\!\!-\!\!1 \Gamma \lambda;1 L) 
W(\kappa' \kappa'\!\!-\!\!1 \Gamma' \kappa;1 L') \\
\\[-6pt]
&\times&\!\!\!\!\!\!
\left[
b^*_{\lambda \mu}b^*_{\kappa'\nu'}
(\!-\!1)^{\mu'} \! b_{\lambda\!-\!\mu'}
(\!-\!1)^{\nu} \! b_{\kappa\!-\!\nu}
\!\!+\!\!
\delta_{\kappa\kappa'} (\!-\!1)^{\nu} \delta_{-\nu,\nu'}
b^*_{\lambda \mu}(\!-\!1)^{\mu'} \! b_{\lambda\!-\!\mu'}
\!\!+\!\!
\delta_{\lambda'\kappa'} (\!-\!1)^{\mu'} \! \delta_{-\mu',\nu'}
b^*_{\lambda \mu}(\!-\!1)^{\nu} b_{\kappa\!-\!\nu}
\right.\\
\\[-6pt]
&&
\left.
\!\!\!\!\!\!\!\!\!\!+
b^*_{\lambda'\mu'} b^*_{\kappa \nu}
(\!-\!1)^{\mu} b_{\lambda\!-\!\mu}
(\!-\!1)^{\nu'} \! b_{\kappa'\!-\!\nu'}
\!\!+\!\!
\delta_{\lambda\lambda'} (\!-\!1)^{\mu} \delta_{-\mu,\mu'}
b^*_{\kappa \nu} (\!-\!1)^{\nu'} \! b_{\kappa'\!-\!\nu'}
\!\!+\!\!
\delta_{\lambda\kappa} (\!-\!1)^{\mu} \delta_{-\mu,\nu}
b^*_{\lambda'\mu'} (\!-\!1)^{\nu'} \! b_{\kappa'\!-\!\nu'} \!
\right] \! ,
\EA
\label{HrotI}
\eeqa\\[-20pt]
\beqa
\!\!\!\!\!\!\!\!\!\!\!\!\!\!
\BA{lll}
&&\!\!\!\!\!\!\!\!\!\!{\mathcal H}_{rot.\mbox{II}}
\!\!=\!\!
{\displaystyle \frac{\hbar^2}{8\r_0}} \!\!
\sum_k \!
\sum_{\lambda\lambda'\kappa\kappa'\mu\mu'\nu\nu'} \!
\sum_{LL'MM'\Gamma\Gamma'\Lambda\Lambda'} \!
\sum_{JK} \!
{\displaystyle \frac{1}{R_0^2}} \!\!
\left(\!\!
{\displaystyle \frac{r}{R_0}} \!\!
\right)^{\!\!\lambda\!+\!\lambda'\!+\!\kappa\!+\!\kappa'\!-\!2} \!\!
(\!-\!1)^{\kappa\!+\!\kappa'} \!
{\displaystyle \frac{(2\lambda'\!\!+\!\!1) \! (2\kappa'\!\!+\!\!1)}{\sqrt{3}}} \\
\\[-8pt]
~~~&\times&\!\!\!\!\!\!
\sqrt{\!
{\displaystyle
\frac{(2\lambda\!\!+\!\!1)(2\lambda'\!\!+\!\!1)}
{4\pi (2L\!\!+\!\!1)}}
}
{\displaystyle
\sqrt{\! 
\frac{\kappa \kappa'(2\kappa'\!\!-\!\!1)
(2\Gamma'\!\!+\!\!1)}
{4\pi}
}
}
\sqrt{\!{\displaystyle
\frac{(2L\!\!+\!\!1)(2L'\!\!+\!\!1)}{4\pi(2J\!\!+\!\!1)}}} \\
\\[-6pt]
&\times&\!\!\!\!\!\!
\langle \lambda 0 \lambda' 0 | L 0 \rangle \!
\langle \kappa\!\!-\!\!1 0 \kappa'\!\!-\!\!1 0 | L' 0 \rangle \!
\langle L 0 L' 0 | J 0 \rangle \!
\langle \lambda \mu \lambda' \mu' | L \!M \rangle \!
(\!-\!1)^{\nu'} \!\!
\langle L' \!M' \kappa' \nu' | \Gamma' \Lambda' \rangle \!
\langle L \!M \!L' \!M' | \!J \!K \rangle \!
Y_{\!J\!K} \\
\\[-6pt]
&\times&\!\!\!\!\!\!
\langle \Gamma' \Lambda' \kappa\!\!-\!\!1 \nu \!\!+\!\!k |1 k \rangle  
W(\kappa' \kappa'\!\!-\!\!1 \Gamma' \kappa\!\!-\!\!1;1 L') \\
\\[-6pt]
&\times&\!\!\!\!\!\!
\left[ 
(\!-\!1)^{\mu} \! b^*_{\lambda \!-\!\mu}
(\!-\!1)^{\mu'} \! b^*_{\lambda\!-\!\mu'}
(\!-\!1)^{\nu} b_{\kappa -\nu}
(\!-\!1)^{\nu'} b_{\kappa'\!-\!\nu'}
\!\!+\!\!
b^*_{\kappa \nu}b^*_{\kappa'\nu'}
b_{\lambda\mu}b_{\lambda'\mu'}
\!\!+\!\!
\delta_{\lambda\kappa} \! \delta_{\mu \nu} \!
b^*_{\kappa' \nu'} b_{\lambda' \nu}
\!\!+\!\!
\delta_{\lambda\kappa'} \! \delta_{\mu \nu'} \!
b^*_{\kappa \nu} b_{\lambda'\mu'}
\right.\\
\\[-6pt]
&&
\left.
\!\!\!\!\!\!\!\!\!\!+
\delta_{\lambda'\kappa} \delta_{\mu' \nu}
b^*_{\kappa' \nu'} b_{\lambda \mu}
\!\!+\!\!
\delta_{\lambda'\kappa'} \delta_{\mu' \nu'}
b^*_{\kappa \nu} b_{\lambda \mu}
\!\!+\!\!
\delta_{\lambda\kappa} \delta_{\mu \nu}
\delta_{\lambda'\kappa'} \delta_{\mu' \nu'}
\!\!+\!\!
\delta_{\lambda' \kappa} \delta_{\mu' \nu}
\delta_{\lambda \kappa'} \delta_{\mu \nu'} 
\right] \! .
\EA
\label{HrotII}
\eeqa\\[-8pt]
The roton Hamiltonian
${\mathcal H}_{rot.}$
consists of normal-ordered
quartic and quadratic terms with respect to roton operators
and constant terms.
On the other hand, the BMM Hamiltonian
has a quadratic form of
surface-phonon operators.
This is a remarkable difference between them.

There exist many multipole degrees of freedom
in the collective coordinates of
the surface oscillations
in the BMM.
As was done in the BMM,
we also pay special attention to collective excitations with
quadrupole degrees of freedom
since such degrees of freedom play
a fundamental role in almost all nuclei.
Then,
in Eqs.
(\ref{angmomentumJk}),
(\ref{HrotI})
and
(\ref{HrotII}),
we restrict to the case of
$\lambda \!=\! \lambda' \!=\! \kappa \!=\! \kappa'$
and they are rewritten 
in the  following forms:

The {\bf rotational} angular momentum
$\widehat{\mathbf J}_{k}\!$
(\ref{angmomentumJk})
at the nuclear surface $(r\!\!=\!\!R_0)$
is expressed as\\[-16pt]
\begin{eqnarray}
\BA{lll}
\widehat{\mathbf J}_{k}
&\!\!\!=\!\!&\!\!
{\displaystyle \frac{5}{2}} \!
\sqrt{{\displaystyle \frac{2\!\cdot\!3\!\cdot\!5(2L'\!+\!1)}{12\pi}}} \!
\sum_{\mu \mu'} \! \sum_{LML'M'}
W( 2 1 L L';2 2)
\langle 2 0 2 0 | L 0 \rangle Y_{LM} \\
\\[-10pt]
&&\times\!
\langle L M 2 \mu | L' M' \rangle \!
\langle L' M' 2 \mu' |1 k \rangle \!
\left[
b^*_{2 \mu'}(-1)^{\mu} b_{2 \!-\!\mu}
\!-\!
b^*_{2 \mu}(-1)^{\mu'} \! b_{2 \!-\!\mu'}
\!-\!
(-1)^{\mu'} \! \delta_{\mu\!-\!\mu'}
\right] ,
\EA
\label{angmomentumJk2}
\end{eqnarray}\\[-10pt]
in which
we pick up only the term with $L\!=\!0$.
Then we have
a simple formula for
$\widehat{\mathbf J}_{k}$
as \\[-18pt]
\begin{eqnarray}
\BA{l}
\widehat{\mathbf J}_{k}
\!=\!
-
{\displaystyle \frac{\sqrt{2\!\cdot\!5}}{4\pi}} \!
\sum_{\mu \mu'} \! 
\langle 2 \mu 2 \mu' |1 k \rangle 
b^*_{2 \mu}(-1)^{\mu'} \! b_{2 \!-\!\mu'} ,
\EA
\label{angmomentumJk3}
\end{eqnarray}\\[-14pt]
which has the same form as the BMM
angular momentum operator except for the minus sign.
It is expressed in terms of the roton operators
without the constant term.\\[-14pt]

The roton Hamiltonian
$
{\mathcal H}_{rot.}
(
\!\equiv\!
{\mathcal H}_{rot.\mbox{I}}
\!+\!
{\mathcal H}_{rot.\mbox{II}}
)
$
(\ref{rot.H})
reduces to simpler forms
in terms of the roton operators 
$b^*_{2 \mu}$
and
$b_{2 \mu}$
as,\\[-14pt]
\beqa
\!\!\!\!\!\!\!\!
\BA{lll}
&&\!\!\!\!\!\!\!\!\!\!{\mathcal H}_{rot.\mbox{I}}
\!\!=\!\!
{\displaystyle \frac{\hbar^2}{8\r_0}} 
{\displaystyle \frac{1}{R_0^2}} 
\left( \!
{\displaystyle \frac{r}{R_0}} \!
\right)^{6} 
{\displaystyle \frac{5\!\cdot\!5}{3}}~\! 
{\displaystyle
\frac{2\!\cdot\!3\!\cdot\!5}{4\pi}
} \!
\sum_k \!
\sum_{\mu\mu'\nu\nu'} \!
\sum_{LL'MM'\Gamma\Gamma'\Lambda\Lambda'} \!
\sum_{JK}(\!-\!1)^{k\!+\!J} \\
\\[-6pt]
&\times&\!\!\!\!\!\!
\sqrt{\!{\displaystyle
\frac{(\!2L\!\!+\!\!1\!) \! (\!2L'\!\!+\!\!1\!) \!
(\!2\Gamma\!\!+\!\!1\!) \! (\!2\Gamma'\!\!+\!\!1\!)}
{4\pi(2J\!\!+\!\!1)}}} \!
\langle 2 0 1 0 | L 0 \rangle \!
\langle 2 0 1 0 | L' 0 \rangle \!
\langle L 0 L' 0 | J 0 \rangle \!
\langle L \!M 2 \mu' | \Gamma \!\Lambda \rangle \!
\langle L' \!M' 2 \nu' | \Gamma' \!\Lambda' \rangle \\
\\[-6pt]
&\times&\!\!\!\!\!\!
\langle L \!M \!L' \!M' | \!J \!K \rangle \!
Y_{\!\!J\!K} \!
\langle \Gamma \Lambda 2 \mu |1 k \rangle \!
\langle \Gamma' \Lambda' 2 \nu |1\!\!-\!\!k \rangle \!
W(2 1 \Gamma 2;1 L) 
W(2 1 \Gamma' 2;1 L') \\
\\[-6pt]
&\times&\!\!\!\!\!\!
\left[
b^*_{2 \mu} b^*_{2 \nu'}
(\!-\!1)^{\mu'} b_{2 \!-\!\mu'}
(\!-\!1)^{\nu} b_{2 \!-\!\nu}
\!+\!
(\!-\!1)^{\nu} \delta_{-\nu,\nu'}
b^*_{2 \mu}(\!-\!1)^{\mu'} b_{2 \!-\!\mu'}
\!\!+\!\!
(\!-\!1)^{\mu'} \! \delta_{-\mu',\nu'}
b^*_{2 \mu}(\!-\!1)^{\nu} b_{2 \!-\!\nu}
\right.\\
\\[-6pt]
&&
\left.
\!\!\!\!\!\!\!\!\!\!+
b^*_{2 \mu'} b^*_{2 \nu}
(\!-\!1)^{\mu} b_{2 \!-\!\mu}
(\!-\!1)^{\nu'} b_{2 \!-\!\nu'}
\!+\!
(\!-\!1)^{\mu} \delta_{-\mu,\mu'}
b^*_{2 \nu} (\!-\!1)^{\nu'} b_{2 \!-\!\nu'}
\!+\!
(\!-\!1)^{\mu} \delta_{-\mu,\nu}
b^*_{2 \mu'} (\!-\!1)^{\nu'} b_{2 \!-\!\nu'} \!
\right] \! ,
\EA
\label{HrotI2}
\eeqa
\vspace{-0.1cm}
\beqa
\!\!\!\!\!\!\!\!\!\!
\BA{lll}
&&\!\!\!\!\!\!\!\!\!\!{\mathcal H}_{rot.\mbox{II}}
\!\!=\!\!
{\displaystyle \frac{\hbar^2}{8\r_0}} 
{\displaystyle \frac{1}{R_0^2}} \!
\left(\!
{\displaystyle \frac{r}{R_0}} \!
\right)^{6} 
{\displaystyle \frac{5\!\cdot\!5}{\sqrt{3}}}~\!
{\displaystyle
\frac{2\!\cdot\!\sqrt{3}\!\cdot\!5}{4\pi}
} \!
\sum_k \!
\sum_{\mu\mu'\nu\nu'} \!
\sum_{LL'MM'\Gamma\Gamma'\Lambda\Lambda'} \!
\sum_{JK} \\
\\[-6pt]
~~~&\times&\!\!\!\!\!\!
\sqrt{\!
{\displaystyle
\frac{2\Gamma'\!\!+\!\!1}{2L\!\!+\!\!1}
}
} \!
\sqrt{\!
{\displaystyle
\frac{(\!2L\!\!+\!\!1\!)(\!2L'\!\!+\!\!1\!)}{4\pi(2J\!\!+\!\!1)}
}
} \!
\langle 2 0 2 0 | L 0 \rangle \!
\langle 1 0 1 0 | L' 0 \rangle \!
\langle L 0 L' 0 | J 0 \rangle \!
\langle 2 \mu 2 \mu' | L \!M \rangle \!
(\!-\!1)^{\nu'} \!\!
\langle L' \!M' 2 \nu' | \Gamma' \Lambda' \rangle \\
\\[-10pt]
&\times&\!\!\!\!\!\!
\langle L \!M \!L' \!M' | \!J \!K \rangle \!
Y_{\!J\!K} 
\langle \Gamma' \Lambda' 1 \nu \!\!+\!\!k |1 k \rangle  
W(2 1 \Gamma' 1;1 L') \\
\\[-6pt]
&\times&\!\!\!\!\!\!
\left[ 
(\!-\!1)^{\mu} b^*_{2 \!-\!\mu}
(\!-\!1)^{\mu'} b^*_{2 \!-\!\mu'}
(\!-\!1)^{\nu} b_{2 -\nu}(\!-\!1)^{\nu'} b_{2 \!-\!\nu'}
\!+\!
b^*_{2 \nu}b^*_{2 \nu'}b_{2 \mu}b_{2 \mu'}
\!+\!
\delta_{\mu \nu} b^*_{2 \nu'} b_{2 \nu}
\!+\!
\delta_{\mu \nu'}
b^*_{2 \nu} b_{2 \mu'}
\right.\\
\\[-6pt]
&&
\left.
\!\!\!\!\!\!\!\!\!\!+
\delta_{\mu' \nu}b^*_{2 \nu'} b_{2 \mu}
\!+\!
\delta_{\mu' \nu'}
b^*_{2 \nu} b_{2 \mu}
\!+\!
\delta_{\mu \nu}\delta_{\mu' \nu'}
\!+\!
\delta_{\mu' \nu}\delta_{\mu \nu'} 
\right] \! .
\EA
\label{HrotII2}
\eeqa
Introducing
$f(r)$
defined as
$
f(r)
\!\equiv\!
{\displaystyle 
\frac{1}{(4\pi)^2}
\frac{\hbar^2}{8\r_0}
\frac{1}{R_0^2}
} \!
\left( \!
{\displaystyle \frac{r}{R_0}} \!
\right)^{6}
$, 
finally we reach the final expression for the roton Hamiltonian
${\mathcal H}_{rot.}$
given in the following form:\\[-14pt]
\beqa
\!\!\!\!
\BA{lll}
&&\!\!\!\!\!\!\!\!\!\!{\mathcal H}_{rot.}
\!=\!
-
{\displaystyle \frac{100}{3}}
f(r) \!
\sum_k \!
\sum_{\nu}  
\langle 2 \!\!-\!\!\nu 1 \nu \!\!+\!\!k |1 \!k \rangle 
\!+\!
{\displaystyle \frac{8}{5}}
f(r) \!
\sum_{\mu} \!
b^*_{2 \mu} b_{2\mu}
\!+\!
{\displaystyle \frac{4}{5}} 
f(r) \!
\sum_{\mu} \!
b^*_{2 \mu} b_{2 \mu} \!
\sum_{\nu} \!
b^*_{2 \nu} b_{2 \nu} \\
\\[-8pt]
&&~\!-
{\displaystyle
\frac{50}{3}
} 
f(r) \!
\sum_k \!
\sum_{\nu} \!
\left\{ \!
\langle 2 \nu 1 k |1 \nu \!\!+\!\!k \rangle
(-1)^{\nu} b^*_{2 -\nu} b_{2 \nu}
\!+\!
3
\langle 2 \!\!-\!\!\nu 1 \nu \!\!+\!\!k |1 \!k \rangle
b^*_{2 \nu} b_{2 \nu}
\right\}  \\
\\[-8pt]
&&~\!+\!
{\displaystyle \frac{27}{40}} 
f(r) \!
\sum_k \!
\sum_{\mu\mu'\nu\nu'} \!\!
\sum_{M\Lambda\Lambda'} \!
(\!-\!1)^{k\!+\!M} \!
\langle 2 \mu 1 M | 1 \Lambda \rangle \!
\langle 2 \mu' 1 \Lambda | 1 k \rangle \!
\langle 2 \nu 1 \!\!-\!\!M | 1 \Lambda' \rangle \!
\langle 2 \nu' 1 \Lambda' | 1 \!\!-\!\!k \rangle \\
\\[-6pt]
&&~\!\times
\left[ 
2
(\!-\!1)^{\nu} \delta_{-\nu,\nu'}
b^*_{2 \mu}(\!-\!1)^{\mu'} b_{2 \!-\!\mu'}
\!+\!
( \! 1 \!+\! (\!\!-\!1\!)^{\Lambda\!+\!\Lambda'} \! ) 
(\!-\!1)^{\mu} \delta_{-\mu,\nu}
b^*_{2 \mu'} (\!-\!1)^{\nu'} \! b_{2 \!-\!\nu'}
\right] \\
\\[-6pt]
&&~
\!-\!
{\displaystyle
\frac{50}{3}
} 
f(r) \!
\sum_k \!
\sum_{\mu \nu} \!
\langle 2 \!\!-\!\!\nu 1 \nu \!\!+\!\!k |1 k \rangle \!\!
\left[ 
b^*_{2 \mu} (\!-\!1)^{\mu} b^*_{2 \!-\!\mu}
b_{2 \nu} (\!-\!1)^{\nu} b_{2 -\nu} 
\!\!+\!\!
b^*_{2 \nu} (\!-\!1)^{\nu} b^*_{2 -\nu}
b_{2 \mu} (\!-\!1)^{\mu} b_{2 -\mu}
\right]  \\
\\[-6pt]
&&~\!+\!
{\displaystyle \frac{27}{40}} 
f(r) \!
\sum_k \!
\sum_{\mu\mu'\nu\nu'} \!\!
\sum_{M\Lambda\Lambda'} \!
(\!-\!1)^{k\!+\!M}
\langle 2 \mu 1 M | 1 \Lambda \rangle \!
\langle 2 \mu' 1 \Lambda | 1 k \rangle
\langle 2 \nu 1 \!\!-\!\!M | 1 \Lambda' \rangle \!
\langle 2 \nu' 1 \Lambda' | 1 \!\!-\!\!k \rangle \\
\\[-4pt]
&&~\!\times 
\left[
b^*_{2 \mu} b^*_{2 \nu'}
(\!-\!1)^{\mu'} b_{2 \!-\!\mu'}
(\!-\!1)^{\nu} b_{2 \!-\!\nu}
\!+\!
b^*_{2 \mu'} b^*_{2 \nu}
(\!-\!1)^{\mu} b_{2 \!-\!\mu}
(\!-\!1)^{\nu'} \! b_{2 \!-\!\nu'} \!
\right]
\!+\!
\cdots ,
\EA
\label{Hrot}
\eeqa\\[-4pt]
where we omit terms
with higher rank of angular momenta
$L, L'$ and $J$ etc.
on the computation of
(\ref{HrotI2})
and
(\ref{HrotII2})
because their explicit expressions become too long to write. 
The roton Hamiltonian at the equilibrium nuclear surface
is given by 
${\mathcal H}_{rot.}|_{r\!=\!R_0}$.

The velocity potential 
$\phi(\vx)$ is also expanded as
$
\phi(\vx)
\!\!=\!\!
\sum_{\mu} \!
\pi_{2\mu} \!
\left( \!{\displaystyle \frac{r}{R_0}} \! \right)^{\!2} \!\!
Y_{2\mu}(\theta,\varphi)
$
where
$\pi_{2\mu}$
and
the collective coordinate
$\alpha_{2\mu}$ given previously
are related to
the BMM boson operator.
To get the fluid velocity $\vv(\!\vx\!)$ at the surface,
we also require a surface boundary condition
$
\left.
{\displaystyle \frac{\partial R(\theta,\varphi)}{\partial t}}
\right|_{r\!=\!R_0}
\!=\!
\left.
v_r
\right|_{r\!=\!R_0}
$. 
Using the fluid velocity $\vv\!$
(\ref{fluidvelocity}),
this condition is given as follows:\\[-12pt]
\beqa
\!\!\!\!\!\!\!\!\!\!\!\!\!\!
\BA{lll}
&&
R_0 \!
\sum_{\mu} \!
\dot{\a}_{2\mu}Y_{2\mu}
\!=\!
-
\left.
\sum_{\mu} \!
\pi_{2\mu} 
{\displaystyle \frac{\partial}{\partial r}} \!\!\!
\left( \! {\displaystyle \frac{r}{R_0}} \! \right)^{\!\!2} \!\!
Y_{2\mu}
\right|_{r\!=\!R_0} \\
\\[-10pt]
&&~\!-\!
{\displaystyle \frac{i\hbar}{2\r_0}} \!
\sum_{\mu\mu'} \!
\left\{ \!
b^*_{2\mu} \!\!
\left( \! {\displaystyle \frac{r}{R_0}} \! \right)^{\!2} \!\!
Y_{2\mu}^*
b_{2\mu'} \!\!
\left.
{\displaystyle \frac{\partial }{\partial r}} \!\!
\left( \! {\displaystyle \frac{r}{R_0}} \! \right)^{\!2} \!\!
Y_{2\mu'}
\right|_{r\!=\!R_0}
\!-\!
b_{2\mu} \!\!
\left( \! {\displaystyle \frac{r}{R_0}} \! \right)^{\!2} \!\!
Y_{2\mu}
b^*_{2\mu'} \!\!
\left.
{\displaystyle \frac{\partial }{\partial r}} \!\!
\left( \! {\displaystyle \frac{r}{R_0}} \! \right)^{\!2} \!\!
Y_{2\mu'}^*
\right|_{r\!=\!R_0} \!
\right\}  \\
\\[-10pt]
&&~\!=\!
-{\displaystyle \frac{2}{R_0}} \!\!
\sum_{\mu} \!
\pi_{2\mu} 
Y_{2\mu} 
\!-\!
{\displaystyle \frac{i\hbar}{2\r_0}} 
{\displaystyle \frac{2}{R_0}} \!\!
\sum_{\mu} \!
\left\{ \!
\left( \!
\sum_{\mu'} 
(-1)^{\mu'} b^*_{2-\mu'} 
Y_{2\mu'} \!
\right) \!
b_{2\mu} 
\!-\!
\left( \!
\sum_{\mu'} 
b_{2\mu'} 
Y_{2\mu'} \!
\right) \!
(-1)^{\mu} b^*_{2-\mu} \!
\right\} \! 
Y_{2\mu} ,
\EA
\label{fluidvelocity2}
\eeqa\\[-8pt]
which gives the relation between
the time derivative
$\dot{\a}_{2\mu}$
and the
$\pi_{2\mu}$
and
the roton operator
$b_{2\mu}$
as \\[-18pt] 
\beqa
\BA{l}
\dot{\a}_{2\mu}
\!=\!
-{\displaystyle \frac{2}{R^{2}_0}} 
\pi_{2\mu}
\!-\!
{\displaystyle \frac{i\hbar}{2\r_0}} 
{\displaystyle \frac{2}{R^{2}_0}} \!
\left\{ \!
\left( \!
\sum_{\mu'}  
(-1)^{\mu'} b^*_{2-\mu'} 
Y_{2\mu'} \!
\right) \!
b_{2\mu} 
\!-\!
\left( \!
\sum_{\mu'} 
b_{2\mu'} 
Y_{2\mu'} \!
\right) \!
(-1)^{\mu} b^*_{2-\mu} \!
\right\} \! .
\EA
\label{fluidvelocity3}
\eeqa\\[-12pt]
It should be noticed that 
in the above there exist bi-linear terms in
$b_{2\mu}$
and
$(-1)^{\mu} b^*_{2-\mu}$.
This is a remarkable aspect
of the relation.
However,
if we discard these terms,
we can reach the well-known result that
the $\pi_{2\mu}$
is the canonical conjugate variable to
the $\a_{2\mu}$.



\newpage

\def\thesection{\arabic{section}}
\setcounter{equation}{0}
\renewcommand{\theequation}{\arabic{section}.\arabic{equation}}
\section{Determination of Clebsch parameters
through a one-form gauge potential}

\vspace{-0.2cm}
 
 Using the equilibrium density $\r_0$,
 the Ziman variables
 $\psi_1$ and $\psi_2$ are expressed
through the two Clebsch parameters $\psi$ and $\lambda$  as\\[-10pt]
\beq
\left.
\begin{array}{ll}
&\psi
=
{\displaystyle \frac{\psi_1}{\psi_2}} ,~
{\psi_1}
=
\psi \psi_2
=
\psi
\sqrt{2 \r_0 \lambda} , \\
\\[-10pt]
&\r_0 \lambda
=
{\displaystyle \frac{\psi_2^2}{2}} ,~
\psi_2
=
\sqrt{2 \r_0 \lambda} .
\end{array}
\right\}
\label{newvariables2}
\eeq\\[-6pt]
Substituting
(\ref{newvariables2})
into the Ziman field operators
$
\Psi
\!=\!
{\displaystyle \frac{1}{\sqrt{2\hbar}}}
(\psi_1\!+\!i\psi_2)
$
and
$
\Psi^*
\!=\!
{\displaystyle \frac{1}{\sqrt{2\hbar}}}
(\psi_1\!-\!i\psi_2)
$,
then we recover
the expression of Allcock-Kuper
\cite{AllKup.55}
for $\Psi$ and $\Psi^*$:\\[-8pt]
\beq
\left.
\begin{array}{ll}
&\Psi
\!=\!
{\displaystyle \sqrt{\frac{\r_0 \lambda}{\hbar}}}
(\psi + i) ,\\
\\[-10pt]
&\Psi^*
\!=\!
{\displaystyle \sqrt{\frac{\r_0 \lambda}{\hbar}}}
(\psi - i) .
\end{array}
\right\}
\label{newfieldoperators2}
\eeq
For the incompressible fluid,
the continuity equation
leads to the condition\\[-16pt]
\beqa
\mbox{div} \vv
\!=\!
-
\nabla^2 \phi
\!+\!
\frac{i\hbar}{2\r_0}
\left( \Psi^* \nabla^2 \Psi \!-\! \Psi \nabla^2 \Psi^* \right)
\!=\!
0 .
\label{condition2}
\eeqa\\[-12pt]
The collective flow is assumed to be {\bf irrotaional},
i.e.,
$
\nabla^2 \phi
\!=\!
0 
$
which means
$
\nabla^2 \Psi
\!=\!
\nabla^2 \Psi^*
\!=\!
0
$
where the Laplacian
in the spherical polar-coordinate is expressed as\\[-8pt]
\beq
\nabla^2
\!=\!
{\displaystyle 
\frac{\partial^2}{\partial r^2}
\!+\!
\frac{2}{r}
\frac{\partial}{\partial r}
\!+\!
\frac{1}{r^2}
\frac{\partial^2}{\partial \theta^2}
\!+\!
\frac{\cos \theta}{r^2 \sin \theta}
\frac{\partial}{\partial \theta}
\!+\!
\frac{1}{r^2 \sin^2 \theta}
\frac{\partial^2}{\partial \varphi^2} .
}
\eeq

According to Jackiw
\cite{Jackiw.02, JackiwPi.00},
the two Clebsch parameters $\psi$ and $\lambda$
are constructed as\\[-10pt]
\beq
\left.
\begin{array}{ll}
&\lambda
\!=\!
2 \!
\left( \!
1
\!-\!
\sin^2 {\displaystyle \frac{f}{2}}
\sin^2 \theta \!
\right) \! ,~
{\displaystyle \frac{\partial \lambda}{\partial \varphi}}
\!=\!
0 , \\
\\[-18pt]
&~~~~~~~~~~~~~~~~~~~~~~~~~~~~~~~~~~~~~~~~~~~~~~
{\displaystyle \frac{\partial \Psi}{\partial \varphi}}
\!=\!
{\displaystyle \sqrt{\frac{\r_0 \lambda}{\hbar}}} , \\
\\[-12pt]
&\psi
\!=\!
\varphi
\!+\!
\tan^{-1} \!
\left( \!
\tan {\displaystyle \frac{f}{2}}
\cos \theta \!
\right) \! ,~
{\displaystyle \frac{\partial \psi}{\partial \varphi}}
\!=\!
1 ,
\end{array}
\right\}
\label{Clebschparametrized}
\eeq
where a profile function
$f  (\!=\!\! f(r))$
will be determined later.
Denoting by differentiation with respect to $r$ by prime ,
the differential formulas for $\psi$ and $\lambda$
with respect to $r$ and $\theta$ are given as\\[-8pt]
\beq
\!\!\!\!\!\!
\left.
\begin{array}{ll}
&
{\displaystyle \frac{\partial \lambda}{\partial r}}
\!=\!
\sin \! f \sin^2 \! \theta f' , \\
\\[-18pt]
&~~~~~~~~~~~~~~~~~~~~~~~~~~~~
{\displaystyle \frac{\partial \Psi}{\partial r}}
\!=\!
-
{\displaystyle \sqrt{\!\frac{\r_0 \lambda}{\hbar}}} \!\!
\left\{ \!
{\displaystyle \frac{1}{2  \lambda}}
\sin \! f \sin^2 \! \theta f'
(\psi \!+\! i)
\!-\!
{\displaystyle
\frac{\cos \! \theta}
{2 \!
\left( \!1 \!-\! \sin^2 \! \frac{f}{2}
\sin^2 \! \theta \!\right)}
} 
f' \!
\right\}, \\
\\[-12pt]
&
{\displaystyle \frac{\partial \psi}{\partial r}}
\!=\!
{\displaystyle
\frac{\cos \! \theta}
{2 \!
\left( \!1 \!-\! \sin^2 \! \frac{f}{2}
\sin^2 \! \theta \!\right)}
} 
f',
\end{array}
\right\}
\label{Clebschparameterdiffr}
\eeq
\beq
\!\!\!\!\!\!
\left.
\begin{array}{ll}
&
{\displaystyle \frac{\partial \lambda}{\partial \theta}}
\!=\!
-2 \sin^2 \! \frac{f}{2} \sin \! 2 \theta  , \\
\\[-22pt]
&~~~~~~~~~~~~~~~~~~~~~~~~~~~~~~~~~
{\displaystyle \frac{\partial \Psi}{\partial \theta}}
\!=\!
-
{\displaystyle \sqrt{\!\frac{\r_0 \lambda}{\hbar}}} \!\!
\left\{ \!
{\displaystyle \frac{1}{ \lambda}}
\sin^2 \! \frac{f}{2} \sin \! 2 \theta 
(\psi \!+\! i)
\!+\!
{\displaystyle
\frac{\sin \! f \sin \! \theta}
{2 \!
\left( \!1 \!-\! \sin^2 \! \frac{f}{2}
\sin^2 \! \theta \!\right)}
} \!
\right\}, \\
\\[-16pt]
&
{\displaystyle \frac{\partial \psi}{\partial \theta}}
\!=\!
-{\displaystyle
\frac{\sin \! f \sin \! \theta}
{2 \!
\left( \!1 \!-\! \sin^2 \! \frac{f}{2}
\sin^2 \! \theta \!\right)}
} .
\end{array}
\right\}
\label{Clebschparameterdifftheta}
\eeq\\[-10pt]
Substitution of
(\ref{Clebschparametrized}),
(\ref{Clebschparameterdiffr})
and
(\ref{Clebschparameterdifftheta})
casts $\nabla^2 \Psi \!=\! 0$ into
the following equation:\\[-6pt]
\beq
\!\!\!\!\!\!\!\!
\begin{array}{ll}
&
\nabla^2 \Psi
\!=\!
-
{\displaystyle \sqrt{\!\frac{\r_0 \lambda}{\hbar}}} \!
\left[ \!
\left\{ \!
{\displaystyle
\frac{1}
{\left\{ \! 2 \!
\left( \!1 \!-\! \sin^2 \!\! \frac{f}{2} \!
\sin^2 \!\! \theta \!\right) \!
\right\}^{\!2}
}
} \!\!
\left( \!
{\displaystyle \frac{1}{4}} \!
\sin^2 \!\! f \! \sin^4 \! \theta \! f'^2
\!+\!
{\displaystyle \frac{4}{r^2}} \!
\sin^4 \!\! \frac{f}{2}
\sin^2 \! \theta \cos^2 \! \theta \!
\right)
\right.
\right. \\
\\[-10pt]
&
\left.
\left.
\!+
{\displaystyle
\frac{1}{2} 
\frac{1}
{2 \!
\left( \!1 \!-\! \sin^2 \!\! \frac{f}{2} \!
\sin^2 \!\! \theta \!\right)
}
} \!\!
\left( \!\!
\cos \! f \! \sin^2 \!\! \theta f'^2
\!\!+\!\!
\sin \! f \! \sin^2 \!\! \theta f''
\!\!+\!\!
2 \! \sin \! f \! \sin^2 \!\! \theta {\displaystyle \frac{1}{r}} f'
\!\!+\!\!
{\displaystyle \frac{4}{r^2}} \!
\sin^4 \!\! \frac{f}{2} \!
\left( 3 \! \cos^2 \! \theta \!\!-\!\! 1\right) \!\!
\right) \!\!
\right\} \!\!
\right.
\left( \! \psi \!\!+\!\! i \!\right) \\
\\[-10pt]
&
-
{\displaystyle
\frac{\cos \! \theta}
{\left\{ \! 2 \!
\left( \!1 \!-\! \sin^2 \!\! \frac{f}{2} \!
\sin^2 \!\! \theta \!\right) \!
\right\}^{\!2}
}
} \!\!
\left\{ \!
\sin \! f \! \sin^2 \!\! \theta f'^2
\!\!+\!
2 \!
\left( \!1 \!-\! \sin^2 \!\! \frac{f}{2} \sin^2 \!\! \theta \right) \!\! f''
\!\!+\!
4 \!
\left( \!1 \!-\! \sin^2 \!\! \frac{f}{2} \sin^2 \!\! \theta \right) \!
{\displaystyle \frac{1}{r}} f'
\!-\!
{\displaystyle \frac{4}{r^2}} \!
\sin \! f \!
\right\} \\
\\[-10pt]
&
\left.
+
{\displaystyle
\frac{\cos \! \theta}
{\left\{ \! 2 \!
\left( \!1 \!-\! \sin^2 \!\! \frac{f}{2} \!
\sin^2 \!\! \theta \!\right) \!
\right\}^{\!2}
}
} \!\!
\left\{ \!
\sin \! f \! \sin^2 \!\! \theta f'^2
\!-\!
{\displaystyle \frac{4}{r^2}}
\sin \! f \!
\sin^2 \!\! \frac{f}{2} \sin^2 \!\! \theta \!
\right\} \!
\right] \\
\\[-10pt]
&
=
-
{\displaystyle \sqrt{\!\frac{\r_0 \lambda}{\hbar}}} \!\!
\left[ \!
\left\{ \!
{\displaystyle
\frac{1}
{\left\{ \! 2 \!
\left( \!1 \!-\! \sin^2 \!\! \frac{f}{2} \!
\sin^2 \!\! \theta \!\right) \!
\right\}^{\!2}
}
} \!
\left( \!\!
\left( \!
\left( \!1 \!-\! \sin^2 \!\! \frac{f}{2} \!
\sin^2 \!\! \theta \!\right)^2
\!\!-
\cos^2 \!\! \theta \!
\right) \!
\! f'^2
\!+\!
{\displaystyle \frac{4}{r^2}} \!
\sin^4 \!\! \frac{f}{2}
\left( \! 1 \!-\! \sin^2 \!\! \frac{f}{2} \! \right) 
\sin^4 \! \theta \!
\right)
\right.
\right. \\
\\[-10pt]
&\!\!\!\!\!\!\!
\left.
\left.
\!+
{\displaystyle 
\frac{\sin^2 \!\! \theta}
{4 \!\!
\left( \!1 \!\!-\!\! \sin^2 \!\! \frac{f}{2} \!
\sin^2 \!\! \theta \!\right)
}
} \!\!\!
\left( \!\!
\sin \! f  \! f''
\!\!+\!\!
2 \! \sin \! f \! {\displaystyle \frac{1}{r}} \! f'
\!\!+\!\!
{\displaystyle
\frac{8}{r^2} \!
\sin^4 \!\! \frac{f}{2} 
 \frac{\cos^2 \! \theta}{\sin^2 \! \theta}
 } \!
 \right) \!\!
\right\} \!\!\!
\right.
\left( \! \psi \!+\! i \!\right) \!
\left.
\!-
{\displaystyle
\frac{\cos \! \theta}
{2 \!\!
\left( \!1 \!\!-\!\! \sin^2 \!\! \frac{f}{2} \!
\sin^2 \!\! \theta \!\right) 
}
} \!\!\!
\left\{ \!\!
f''
\!\!+\!\!
{\displaystyle \frac{2}{r}} \! f'
\!\!-\!\!
{\displaystyle \frac{2}{r^2}} \!
\sin \! f \!\!
\right\} \!\!
\right]  \\
\\[-10pt]
&
=
-
{\displaystyle \sqrt{\!\frac{\r_0 \lambda}{4 \hbar}}} \!\!
\left[ \!
\left\{ \!\!
{\displaystyle
\frac{2 \! \sin^4 \!\! \frac{f}{2} \!
\left( \! 1 \!-\! \sin^2 \!\! \frac{f}{2} \! \right) \! \sin^4 \! \theta}
{r^2 \cos^2 \! \theta }
}   
\!+\!
{\displaystyle 
\frac{\sin \! f  \! \sin^2 \!\! \theta}
{2 \! \cos \! \theta}
} \!\!
\left( \!\!
f''
\!\!+\!\!
{\displaystyle \frac{2}{r}} f' \!\! 
\right) \!
\right\} \!
\left( \! \psi \!+\! i \!\right) 
\!-\!
\left( \!\!
f''
\!\!+\!
{\displaystyle \frac{2}{r}} f'
\!\!-\!\!
{\displaystyle \frac{2}{r^2}} \!
\sin \! f \!\!
\right) \!
\right] \!
\!=\!
0 . 
\end{array}
\label{ClebschparameterLaplacian}
\eeq\\[-6pt]
In the above,
to eliminate the term $f'^2$,
we have assumed the auxiliary condition\\[-10pt]
\beq
1 - \sin^2 \!\! \frac{f}{2} \sin^2 \! \theta
\!=\!
\cos \! \theta ,
\label{auxicondisol}
\eeq\\[-14pt]
which implies that either $\cos \! \theta \!=\! 1$ or
$\cos \! \theta \!=\! \cot^2 \!\! {\displaystyle \frac{f}{2}}$.
This condition plays an essential role
in the solution of the Laplace equation $\nabla^2 \Psi \!=\! 0$.
For the present central purpose,
from the outset,
we demand that
the profile function
$f(=\! f(r))$
obeys the differential equation
$
f''
\!+\!
{\displaystyle \frac{2}{r}} f'
\!-\!
{\displaystyle \frac{2}{r^2}} 
\sin \! f
\!=\!
0 
$.
This equation to determine $f$ was previously obtained
by Jackiw-Pi
\cite{JackiwPi.00}.
It has been integrated numerically by Bergner
 ($\!$Ref.[10] $\!$of$\!$ \cite{JackiwPi.00}$\!$).
Introducing the dimensionless variable $x$
as $x \!=\! {\displaystyle \frac{r}{R_0}}$,
the positive solution is plotted in Fig.1 of 
\cite{JackiwPi.00}.
$\!$They present the solution of $\!f (x)\!$
that is regular at the origin,
$x \!=\! 0$, vanishing linearly with $x(0\!\!<\!\!x\!\!<\!\!1)$,
and tending to $\pi$ in an oscillatory manner for larger $x$.
In the region $0\!\!<\!\!x\!\!<\!\!1$, we have
$\sin \! f  \!\!\approx\!\! 0$ and 
due to  
(\ref{auxicondisol})
$\cos \! \theta \!\!=\!\! 1$.
The coefficient of the term
$\left( \! \psi \!\!+\!\! i \!\right)$
in the last line of equation
(\ref{ClebschparameterLaplacian})
vanishes exactly.
Thus we obtain an almost complete solution
for $\nabla^2 \Psi \!\!=\!\! 0$
over the whole region of $x$.

By projecting onto a fixed direction
$\!\hat{n}^a (\mbox{constant unit vector};a \!\!=\!\! 1\!\!\sim\!\!3)\!$
in the isospin space
and using $U^{-1} dU$,
consider a Clebsch-parameterized gauge potential 1-form $a$
\cite{Jackiw.02, JackiwPi.00} 
given as \\[-10pt]
\beq
a
\!=\!
i \mbox{Tr} \hat{n}^a \! \sigma^a U^{-1} dU ,
(\mbox{using the summation convention in}~\hat{n}^a \! \sigma^a) ,
\label{gaugepotA}
\eeq\\[-10pt]
where
$
U
\!=\!
e^{{\displaystyle \hat{\omega}^a \! \frac{\sigma^a}{2i} \! f}} \!
\!=\!
\cos  \! {\displaystyle \frac{f}{2}}
\!-\!
i \sigma^a \! \hat{\omega}^a \! \sin \! {\displaystyle \frac{f}{2}} ,~
\hat{\omega}^a
\!\equiv\!
(\sin \! \theta \! \cos \! \varphi, \sin \! \theta \! \sin \! \varphi, \cos \! \theta)
$
and $\sigma^a$
is the Pauli matrices.
Taking $\hat{n}^a$ to point in the third direction,
the 1-form $a$
(\ref{gaugepotA})
is expressed as\\[-10pt]
\beq
a
\!=\!
\cos \theta df
\!-\!
\sin f
\sin \theta d \theta
\!-\!
(1 \!-\! \cos f) \sin^2 \! \theta d \varphi ,~
(\mbox{see  Appendix A}) .
\label{gaugepotA2}
\eeq\\[-10pt]
The integral
$\!{\displaystyle \frac{1}{24 \pi^{\!2}} \!\!\! \int} \!\!\mbox{Tr}
(\!U^{\!-1} \! dU\!)^{\!3}\!$
is an integer, $\!$i.e.,$\!$ 
the \underline{winding number} $\!\alpha_{\!i,j}\!$
or the \underline{quantized helicity} 
\cite{KuzMikh.80,JackiwPi.00} 
related to the Chern-Simons numbers of
a non-Abelian vacuum gauge potential
\cite{DeserJackiwTempleton.82}.
For the motion of quantum vortices,
Nambu set up an Hamilton-Jacobi formalism 
and reached the conclusion that
the corresponding Hamilton-Jacobi functions
are
the Clebsch potentials
\cite{Nambu.80}.
On the other hand, the canonical quantization 
in terms of the Clebsch parameters
has been developed by Rasetti-Regge.
Their commutation relations are those of a current algebra 
\cite{RasettiRegge.75,RasettiRegge.84,RasettiRegge.85}.


\newpage

\def\thesection{\arabic{section}}
\setcounter{equation}{0}
\renewcommand{\theequation}{\arabic{section}.\arabic{equation}}
\section{Discussions and further outlook}

In the preceding sections,
we have attempted to formulate a description of the {\bf rotational} velocity field
of the nuclear fluid
through the Clebsch transformation.
In the framework of quantum fluid dynamics,
we have derived
the {\bf vortex} Hamiltonian of the fluid
in terms of the roton operators.
According to the previous considerations,
the quantum fluid-dynamical approach
may be applied
to the three-dimensional nuclear fluid.
We expect that such an application to nuclei will provide
an interesting description of a new kind of elementary energy excitation,
namely, "{\bf vortex modes}"
because such an approach is designed to take
into account essential many-body effects, which were not considered
in the traditional treatment of the rotational collective motion.

On the other hand,
extending Tomonaga's idea
and using Sunakawa's method,
one of the present authors (S.N.)
has developed the collective description of nuclear surface oscillations
and the collective theory of two-dimensional nuclei,
in the context of the first quantized language,
as opposed to the second quantized approach
adopted in the Sunakawa's method.
Applying the Tomonaga's revolutionary idea and
the Sunakawa's discrete integral equation method
for collective theory,
we have developed successfully
an $exact$ canonical momenta approach
to one-dimensional neutron-proton systems 
and a velocity operator approach
to three-dimensional neutron-proton systems
\cite{NishProvi2.16}.
Particularly in the latter approach,
we, however, have restricted the Hilbert space to a subspace
$| \! >$
in which the {\bf vortex operator},
$\mbox{rot}\vv(\vx)$,
satisfies
$\mbox{rot}\vv(\vx)| \! > =  0$,
where
the velocity operator
$\vv(\vx)$ is given as
$
\vv(\vx)
=
- \nabla \! \phi(\vx)
+
\lambda(\vx) \nabla \! \psi(\vx) ,
$
i.e., equation
(\ref{Clebschrep}).

To describe the {\bf rotational} velocity field
of the nuclear fluid ,
we have introduced the Clebsch transformation.
However,
we did not yet fully clarify the {\bf vortex motion}.
It is still an important  problem to be solved.
We present some clues for the solution of such problems:

Firstly, the vorticity ${\mathbf w}$
(\ref{vorticity})
is expressed as
$
{\displaystyle \frac{i\hbar}{2 \r}} \!
\left( \! 
\nabla\Psi \!\!\times\!\! \nabla\Psi^* \!\!-\!\! \nabla\Psi^* \!\!\times\!\! \nabla\Psi \!
\right)
$. 
Since ${\mathbf w}$  is orthogonal with
$\nabla\lambda$ and $\nabla\psi$,
if we add a condition
$
(\nabla\lambda) \!\cdot\! (\nabla\psi)
\!=\!
0
$,
the three vectors
${\mathbf w},\nabla\lambda$ and $\nabla\psi$
are orthogonal to each other.
Under the condition, 
the absolute value of
${\mathbf w}$,
$|{\mathbf w}|$
becomes maximum.
Due to  the Ziman transformation
applied to the complex field operators
$\Psi$ and $\Psi^*$,
that condition
is rewritten as
$
{\displaystyle \frac{i\hbar}{\r}} \!
\left(\nabla\Psi \!\!-\!\! \nabla\Psi^*\right)
\!\cdot\!
{\displaystyle \frac{\Psi\nabla\Psi^* \!\!-\!\! \Psi^*\nabla\Psi}{\Psi \!\!-\!\! \Psi^*}}
$. 
Then,
$
(\nabla\lambda) \!\cdot\! (\nabla\psi)
\!=\!
0
$
is changed to\\[-10pt]
\beqa
\BA{lll}
&&(\nabla\Psi \!-\! \nabla\Psi^*)
\!\cdot\!
(\Psi\nabla\Psi^* \!-\! \Psi^*\nabla\Psi)
\!=\!
0  \\
\\[-4pt]
&&
\!=\!
(\nabla\Psi)
\!\cdot\!
\Psi\nabla\Psi^*
\!-\! 
(\nabla\Psi^*)
\!\cdot\!
\Psi\nabla\Psi^*
\!-\!
(\nabla\Psi)
\!\cdot\!
\Psi^*\nabla\Psi
\!+\!
(\nabla\Psi^*)
\!\cdot\!
\Psi^*\nabla\Psi .
\EA
\label{orthoCondition3}
\eeqa
The first term
$
(\!\nabla\!\Psi\!)
\!\cdot\!
\Psi\!\nabla\!\Psi^* 
(\!=\! \Xi)
$
vanishes,
since $\langle 2 0 2 0 | 1 0 \rangle \!=\! 0$.
Thus $\Xi$
has a form given below\\[-22pt]
\begin{eqnarray}
\!\!\!\!\!\!\!\!\!\!\!\!
\BA{lll}
&&
\Xi 
\!\propto\! 
\langle 2 0 2 0 | 1 0 \rangle 
R_0 \!
\sum_{\mu \mu' \mu''k} 
\langle 2 \mu 1 k| 1 \mu \!\!+\!\! k \rangle \!
\langle 1 \mu \!\!+\!\! k 2 \mu' | 1 \mu'' \!\!+\!\! k \rangle\!
\langle 2 \!\!-\!\! \mu'' 1 \!\!-\!\! k | 1 \!\!-\!\! \mu'' \!\!-\!\! k \rangle
b_{2 \mu}b_{2 \mu'}b^*_{2 \mu''}
\!=\!
0 .
\EA
\label{orthoCondition4}
\end{eqnarray}\\[-22pt]
The other terms also vanish.
Then the above orthogonal condition
is automatically satisfied.
Eckart discussed
an extension of the vorticity $\!w\!$
to
$
\nu (\nabla\lambda) \!\!\times\!\! (\nabla\psi)
$,
where $\nu$ is some scalar function.
$\!$The parameter $\!\nu\!$,
however,
can be taken equal to one without loss of generality
\cite{Eckart.60}.

Next, consider the continuity equation,
$\!
\dot{\r} \!+\! \mbox{div}(\r \vv)
\!=\!
\dot{\r} \!+\! (\vv \!\cdot\! \nabla \! \r) \!+\! \r \mbox{div}\vv
\!=\!
0
$,
in which
$\mbox{div}\vv \!=\! 0$
does not imply that both
$\dot{\r} \!=\! 0$ and $\nabla \r \!=\! 0$.
If the Lagrange differentiation for density satisfies
$
{\displaystyle \frac{D\r}{Dt}}
\!\equiv\! 
\dot{\r} \!+\! (\vv \!\cdot\! \nabla) \r
\!=\!
0 
$,
it leads to
$\mbox{div} \vv \!=\! 0$.
In the BMM,
the expression for
the velocity field $\vv(\vx,t) \!=\! -\nabla \phi(\vx,t)$
plays a central role.
Adding to this,
the collective coordinates
are expanded
around the nuclear equilibrium
and then
the expansion coefficient of $\phi$ reduces to
$\dot{\alpha}$, i.e., $\pi$.
To go from classical fluid dynamics to quantum fluid dynamics
by quantization,
through
$\vv \!=\! -\nabla \! \phi$,
what is essential is
a canonical commutation relation for
$\phi$ and $\r$,
$
[\phi(\vx),\r(\vx')]
\!=\!
i\hbar \delta(\vx \!-\! \vx')
$.
Since we assume
a constant density, $\r \!=\! \r_0$,
we can not see an apparent role for
the above commutation relation.
To make further development of the BMM, 
we naturally arrive at the idea of
the expansion of $\r$ around $\r_0$, i.e.,
$\r \!=\! \r_0 + \r^\prime$ 
\cite{Yee.69,Krothell52,Ziman53,Thell53.56}.\\[-16pt]

Lastly, we prove the gauge invariance of the velocity operator $\vv$
expressed in terms of the Clebsch parameters as
$
\vv
\!\!=\!\!
- \nabla \phi
\!-\!
{\displaystyle \frac{1}{\r}} \pi \nabla \psi ,
$
i.e.,
(\ref{Clebschrep})
where we have used the relation
$\pi \!\!=\!\! -\r \lambda$.\\[-6pt]
We consider the following gauge transformation
with the generating function 
$
\omega \!
\left(
\!=\! 
\omega \! \left(\!{\displaystyle \frac{\pi'}{\r}}, \psi \!\right) \!\!
\right)
$:\\[-16pt]
\beqa
\BA{l}
\!\!\!\!\!\!\!\!
\phi'
\!=\!
\phi
\!+\!
\pi'
{\displaystyle \frac{\partial \omega}{\partial \pi'}}
\!-\!
\omega ,~~
\pi'
\!=\!
\pi
\!+\!
\r
{\displaystyle \frac{\partial \omega}{\partial \psi}} ,~~
\psi'
\!=\!
\psi
\!-\!
\r
{\displaystyle \frac{\partial \omega}{\partial \pi'}} , ~~
\r {\displaystyle \frac{\partial \omega}{\partial \r}}
\!+\!
\omega
\!+\!
\pi' {\displaystyle \frac{\partial \omega}{\partial \pi'}}
\!-\!
\omega
\!=\!
0 .
\EA
\label{gaugetrans}
\eeqa\\[-16pt]
This kind of the gauge transformation
was proposed by Ito
\cite{Ito53,Ito55}.
$\!$Kambe also has formulated a variation of ideal fluid flows 
according to the gauge principle in the Clebsch solution
\cite{Kambe08}.
Substituting the gauge transformation
(\ref{gaugetrans})
into
$
\vv'
\!\!=\!\!
- \nabla \! \phi'
\!\!-\!\!
{\displaystyle \frac{1}{\r}} \pi' \nabla \! \psi' ,
$
we obtain\\[-16pt]
\beqa
\BA{l}
\vv'
\!=\!
- \nabla \!
\left( \!
\phi
\!+\!
\pi'
{\displaystyle \frac{\partial \omega}{\partial \pi'}}
\!-\!
\omega \!
\right)
\!-\!
{\displaystyle \frac{1}{\r}} \!
\left( \!
\pi
\!+\!
\r
{\displaystyle \frac{\partial \omega}{\partial \psi}} \!
\right) \!
\nabla \!
\left( \!
\psi
\!-\!
\r
{\displaystyle \frac{\partial \omega}{\partial \pi'}} \!
\right)
\!=\!
- \nabla \! \phi
\!-\!
{\displaystyle \frac{1}{\r}} \pi \nabla \! \psi 
\!=\!
\vv ,
\EA
\label{gaugetrans2}
\eeqa\\[-14pt]
where we have used
the second and last relations of
(\ref{gaugetrans})
and
the gradient formula for $\omega$
\cite{Ito55}, \\[-18pt]
\beqa
\nabla \omega
\!=\!
{\displaystyle \frac{\partial \omega}{\partial \psi}} \! \nabla \psi
\!+\!
{\displaystyle \frac{\partial \omega}{\partial \pi'}} \! \nabla \pi'
\!-\!
{\displaystyle \frac{1}{\r} \pi'
\frac{\partial \omega}{\partial \pi'}} \! \nabla \r .
\label{gradformula}
\eeqa\\[-16pt]
Thus we prove the gauge invariance of the velocity
$\vv$.
Following
\cite{Ito55},
let us separate $\vv$ as
$\vv \!=\! \vv_0 \!+\! \vv_1$.
Putting
(\ref{gaugetrans})
into the
velocity $\vv'_0$
$\! 
\left( 
\!=\!
- \nabla \! \phi'
\!-\!
{\displaystyle \frac{1}{\r_0}} \pi' \nabla \! \psi' \!
\right) 
\!$
and using
(\ref{gradformula}),
we obtain\\[-10pt]
\beqa
\BA{ll}
\vv'_0
&\!\!\!
\!=\!
- \nabla \!
\left( \!
\phi
\!+\!
\pi'
{\displaystyle \frac{\partial \omega}{\partial \pi'}}
\!-\!
\omega \!
\right)
\!-\!
{\displaystyle \frac{1}{\r_0}} \!
\left( \!
\pi
\!+\!
\r
{\displaystyle \frac{\partial \omega}{\partial \psi}} \!
\right) \!
\nabla \!
\left( \!
\psi
\!-\!
\r
{\displaystyle \frac{\partial \omega}{\partial \pi'}} \!
\right) \\
\\[-8pt]
&
\!=\!
- \nabla \phi
\!-\!
{\displaystyle \frac{1}{\r_0}} \pi \nabla \psi  
\!-\!
{\displaystyle \frac{\r'}{\r_0}} \!
\left( \! 
{\displaystyle \frac{\partial \omega}{\partial \psi}} \! \nabla \psi
\!-\!
\pi' \nabla 
{\displaystyle \frac{\partial \omega}{\partial \pi'}}
\!-\!
{\displaystyle \frac{1}{\r} \pi'
\frac{\partial \omega}{\partial \pi'}} \! \nabla \r \!
\right) \\
\\[-10pt]
&
\!=\!
- \nabla \phi
\!-\!
{\displaystyle \frac{1}{\r_0}} \pi \nabla \psi  
\!-\!
{\displaystyle \frac{\r' }{\r_0}} 
\nabla \!
\left( \! 
\omega
\!-\!
\pi' \nabla 
{\displaystyle \frac{\partial \omega}{\partial \pi'}} \!
\right) \!
\!=\!
\vv_0
\!-\!
{\displaystyle \frac{\r' }{\r_0}} 
\nabla \! \left( \phi \!-\! \phi' \right) ,
\EA
\label{gaugetrans3}
\eeqa\\[-6pt]
where linearization of
$\!{\displaystyle \frac{1}{\r}}\!$
is made as
$\!
{\displaystyle \frac{1}{\r}}
\!\!\approx\!\!
{\displaystyle \frac{1}{\r_0}} \!\!
\left( \!\!
1
\!\!-\!\!
{\displaystyle \frac{\r' }{\r_0}}
\!\!+\!\!
{\displaystyle \frac{\r'^2 }{\r_0^2}} \!
\cdots \!\!
\right)
\!\!$
and the first relation of$\!$
(\ref{gaugetrans})$\!$
is used.
The other velocity component $\vv_1$ is expressed as
$
\vv_1
\!\!=\!\!
{\displaystyle \frac{\r' }{\r_0}} \!
\left( \!
{\displaystyle \frac{1}{\r_0}}
\!+\! 
\cdots \!
\right) \!
\pi \nabla \psi
$.
Since the gauge invariance of the velocity
$\vv$ is guaranteed,
the velocoties $\vv_0$ and $\vv_1$
must be invariant, respectively.
From the right-hand side in the last line of Eq.
(\ref{gaugetrans3}),
we must demand $\nabla \! \left( \phi \!-\! \phi' \right) \!=\! 0$ which means
$
\nabla \omega
\!-\! 
\nabla \pi'
{\displaystyle \frac{\partial \omega}{\partial \pi'}}
\!-\!
\pi'
\nabla {\displaystyle \frac{\partial \omega}{\partial \pi'}}
 \!=\!
0
$.
Further, using this relation and
(\ref{gradformula}),
$\pi' \nabla \psi'$
is calculated as 
$
\pi' \nabla \! \psi' 
\!=\!
\left( \!
\pi
\!\!+\!\!
\r {\displaystyle \frac{\partial \omega}{\partial \psi}} \!
\right) \!
\! \nabla \!\!
\left( \!
\psi
\!\!-\!\!
\r {\displaystyle \frac{\partial \omega}{\partial \pi'}} \!
\right)
\!=\!
\pi \nabla \! \psi
\!+\!
{\displaystyle \frac{\r}{\pi'}} \!\!
\left( \!
\pi'
\!\!-\!\!
\pi
\!\!-\!\!
\r {\displaystyle \frac{\partial \omega}{\partial \psi}} \!\!
\right) \!\!
{\displaystyle \frac{\partial \omega}{\partial \psi}}
\nabla \! \psi
\!=\!
\pi \nabla \psi
$.
Then,
$\vv_1' \!\!=\!\! \vv_1$
is proved.
Thus, we can also prove the gauge invariance of the velocity
$\vv$,
even if we make
the separation of $\vv$ as
$\vv \!=\! \vv_0 \!+\! \vv_1$
under the linearization of
${\displaystyle \frac{1}{\r}}$
around
${\displaystyle \frac{1}{\r_0}}$
as given in the above linearization procedure.

In contrast to the present approach to the vortex motion in nuclei,
we notice the papers
\cite{HolSch.78,HolEck.77}
in which
Holtzwarth, Sch\"{u}tte and Eckart
have derived fluid-dynamical equations of motion
allowing for velocity fields with vorticity.
They have parametrized the amplitude and the phase
of a many-body wave function
of a fermion system
and have given a time-dependent variational derivation
of nuclear fluid dynamics.
It is may be interesting to compare
the approaches of Holztzwarth and collaborators
and of Ziman
for deeper understanding of
nuclear fluid dynamics.


\newpage

\noindent
\centerline{\bf Acknowledgements}

\vspace{0.5cm}

One of the authors (S.N.) would like to
express his sincere thanks to
Professor Constan\c{c}a Provid\^{e}ncia for
kind and warm hospitality extended to him at
the Centro de F\'\i sica, Universidade de Coimbra.
This work was supported by FCT (Portugal) under the project
CERN/FP/83505/2008.
A part of calculations in the early stage of this work
was carried out with the aid of
Y. Ishihara of Kochi University.
The authors thank the Yukawa Institute for Theoretical Physics
at Kyoto University. Discussions during the workshop
YITP-W-16-05 on ``Strings and Fields 2016''
are useful to complete this work.


\newpage

\leftline{\large{\bf Appendix}} 
\appendix

\vspace{-0.5cm}

\def\thesection{\Alph{section}}
\setcounter{equation}{0}  
\renewcommand{\theequation}{\Alph{section}. \arabic{equation}}
\section{Detailed derivation of Eq. (\ref{gaugepotA2})}

~~Using
$
U
\!=\!
\cos  \! \frac{f}{2}
\!-\!
i \sigma^a \! \hat{\omega}^a \! \sin \! \frac{f}{2} ,~
\hat{\omega}^a
\!\equiv\!
(\!\sin \! \theta \! \cos \! \varphi, \sin \! \theta \! \sin \! \varphi, \cos \! \theta\!)
$
given in the previous Section 5,
$dU$ is calculated as\\[-14pt]
\beqa
\left.
\BA{ll}
&dU
\!=\!
-{\displaystyle \frac{1}{2}} \!
\left( \!
\sin \! {\displaystyle \frac{f}{2}}
\!+\!
i \sigma^a \! \hat{\omega}^a \!
\cos \! {\displaystyle \frac{f}{2}} \!
\right) \!
df
\!-\!
i \sigma^a \!
\sin \! {\displaystyle \frac{f}{2}}
d \hat{\omega}^a \! ,\\
\\[-12pt]
&d \hat{\omega}^a
\!=\!
(\cos \! \theta d \theta \cos \! \varphi
\!-\!
\sin \! \theta \! \sin \! \varphi d \! \varphi, 
\cos \! \theta d \theta \sin \! \varphi
\!+\!
\sin \! \theta \! \cos \! \varphi d \! \varphi,
-\sin \! \theta d \! \theta) .
\EA
\right\}
\label{dU}
\eeqa\\[-14pt]
Further using
$U^{-1} \!=\! U^{\dag}$
and
(\ref{dU}),
$
U^{-1}dU
$
is computed as \\[-16pt]
\beqa
\BA{ll}
&
U^{-1}dU
\!=\!
\left( \!
\cos \! {\displaystyle \frac{f}{2}}
\!+\!
i \sigma^a \! \hat{\omega}^a \!
\sin \! {\displaystyle \frac{f}{2}} \!
\right) \!\! 
\left\{ \!
-{\displaystyle \frac{1}{2}} \!
\left( \!
\sin \! {\displaystyle \frac{f}{2}}
\!+\!
i \sigma^a \! \hat{\omega}^a \!
\cos \! {\displaystyle \frac{f}{2}} \!
\right) \!
df
\!-\!
i \sigma^b \!
\sin \! {\displaystyle \frac{f}{2}}
d \hat{\omega}^b \! 
\right\} \\
\\[-12pt]
&
\!=\!
-{\displaystyle \frac{1}{4}} \!
\left( \!
\sin \! f
\!+\!
2 i \sigma^b \hat{\omega}^b
\!-\!
\sigma^b \sigma^c
\hat{\omega}^b \hat{\omega}^c 
\sin \! f \!
\right) \!
df
\!-\!
i{\displaystyle \frac{1}{2}}
 \sigma^b \sin \! f \hat{\omega}^b 
\!+\!
{\displaystyle \frac{1}{2}}
\sigma^b \hat{\omega}^b \sigma^c \!
\left( \! 1 \!-\! \cos \! f \! \right) 
d \hat{\omega}^c .
\EA
\label{U-1dU}
\eeqa\\[-12pt]
We are now in the stage to compute
$
a
\!=\!
i \mbox{Tr} \hat{n}^a \! \sigma^a U^{-1}dU 
$
explicitly.
The trace formulas
$\mbox{Tr} \sigma^a \!=\! 0$
and
$\mbox{Tr} (\sigma^a \! \sigma^b ) \!=\! 2 \delta_{ab}$
are useful. 
For our aim, let us prepare
the following trace formulas:\\[-16pt]
\beqa
\BA{ll}
&
-{\displaystyle \frac{1}{4}} 
i \mbox{Tr} \hat{n}^a \sigma^a \!
\left( \!
\sin \! f
\!+\!
2 i \sigma^b \hat{\omega}^b
\right) \!
df
\!=\!
{\displaystyle \frac{1}{2}}
\mbox{Tr} (\sigma^a \! \sigma^b )
\hat{n}^a \hat{\omega}^b
df
\!=\!
\delta_{ab}
\hat{n}^a \hat{\omega}^b
df
\!=\!
\hat{\omega}^3
df
\!=\!
\cos \! \theta df , \\
\\[-2pt]
&
{\displaystyle \frac{1}{2}}
\mbox{Tr} \hat{n}^a \sigma^a
 \sigma^b \sin \! f
d\hat{\omega}^b
\!=\!
{\displaystyle \frac{1}{2}}
\mbox{Tr} (\sigma^a \! \sigma^b )
\hat{n}^a \sin \! f d \hat{\omega}^b
\!=\!
\delta_{ab}
\hat{n}^a \sin \! f d \hat{\omega}^b
\!=\!
\sin \! f \hat{\omega}^3
\!=\!
- \sin \! f \! \sin \! \theta
d  \theta, \\
\\[-2pt]
&
\mbox{Tr} ( \sigma^a \sigma^b \sigma^c)
\hat{n}^a 
\hat{\omega}^b \hat{\omega}^c 
\!=\!
\mbox{Tr} ( \sigma^a \delta^{bc} \!+\! \sigma^a i \epsilon^{bcd} \sigma^b \sigma^c)
\hat{n}^a 
\hat{\omega}^b \hat{\omega}^c 
\!=\!
i \epsilon^{bcd} \mbox{Tr} ( \sigma^a \sigma^d)
\hat{n}^a 
\hat{\omega}^b \hat{\omega}^c \\
\\[-10pt]
&~~~~~~~~~~~~~~~~~~~~~~~~\!
\!=\!
2 i \epsilon^{bca}
\hat{n}^a 
\hat{\omega}^b \hat{\omega}^c 
\!=\!
2 i \epsilon^{bc3}
\hat{\omega}^b \hat{\omega}^c
\!=\!
2 i \!
\left( \!
\hat{\omega}^1 \hat{\omega}^2
\!-\!
\hat{\omega}^2 \hat{\omega}^1 \!
\right)
\!=\!
0 , \\
\\[-2pt]
& 
{\displaystyle \frac{1}{2}}
i
\mbox{Tr} \hat{n}^a \sigma^a
\sigma^b \hat{\omega}^b \sigma^c \!
\left( \! 1 \!\!-\!\! \cos \! f \! \right) \!
d \hat{\omega}^c 
\!=\!
\left( \! 1 \!\!-\!\! \cos \! f \! \right) \!
{\displaystyle \frac{1}{2}}
i
\!\cdot\!
2 i \epsilon^{bca}
\hat{n}^a \hat{\omega}^b
d \hat{\omega}^c
\!=\!
\left( \! 1 \!\!-\!\! \cos \! f \! \right) \!
{\displaystyle \frac{1}{2}}
i
\!\cdot\!
2 i \epsilon^{bc3}
\hat{\omega}^b 
d \hat{\omega}^c \\
\\[-10pt]
&~~~~~~~~~~~~~~~~~~~~~~~~~~~~~~~~~~~~~\!\!
\!=\!
\left( \! 1 \!\!-\!\! \cos \! f \! \right) \!
{\displaystyle \frac{1}{2}}
i \!
\left\{ 
2 (-i) (-1) \!
\left(
\hat{\omega}^1 d \hat{\omega}^2
\!-\!
\hat{\omega}^2 d \hat{\omega}^1 
\right) 
\right\} \\
\\[-10pt]
&~~~~~~~~~~~~~~~~~~~~~~~~~~~~~~~~~~~~~\!\!
\!=\!
\left( \! 1 \!\!-\!\! \cos \! f \! \right) \!\!
\left\{ \!
- \! \sin \! \theta \! \cos \! \varphi \!
\left( \! \cos \! \theta d \theta \! \sin \! \varphi \! \right)
\!\!-\!\!
\sin \! \theta \! \cos \! \varphi \!
\left( \! \sin \! \theta \! \cos \! \varphi d \! \varphi \! \right)
\right. \\
\\[-10pt]
&~~~~~~~~~~~~~~~~~~~~~~~~~~~~~~~~~~~~~~~~~~\!\!\!
\left.
\!+\!
\sin \! \theta \! \sin \! \varphi \!
\left( \! \cos \! \theta d \theta \cos \! \varphi \! \right) 
\!+\!
\sin \! \theta \! \sin \! \varphi \!
\left( \! -\sin \! \theta \sin \! \varphi d \! \varphi \! \right) 
\right\} \\
\\[-10pt]
&~~~~~~~~~~~~~~~~~~~~~~~~~~~~~~~~~~~~~\!\!
\!=\!
\left(1 - \cos \! f \right) 
\sin^2 \! \varphi d \varphi ,
\EA
\label{U-1dU2}
\eeqa\\[-12pt]
in which we take $\hat{n}^a$ to point in the third direction.
Gathering the above all formulas of
(\ref{U-1dU2}),
at last
we acquire the desired expression for $a$
(\ref{gaugepotA2})
as\\[-6pt]
\beq
a
\!=\!
\cos \theta df
\!-\!
\sin f
\sin \theta d \theta
\!-\!
(1 \!-\! \cos f) \sin^2 \! \theta d \varphi .
\label{gaugepotA3}
\eeq
As described by Jackiw-Pi
\cite{JackiwPi.00},
another formula for
(\ref{gaugepotA3})
in the Clebsch representation for the velocity field
$
\vv,~ 
\vv
\!=\!
-\nabla\phi
+
\lambda \nabla\psi 
$,
is given by
\beq
a
\!=\!
d (- 2 \varphi)
\!+\!
2 \!
\left( \!
1
\!-\!
\sin^2 \! {\displaystyle \frac{f}{2}}
\sin^2 \theta \!
\right) \!
d \!
\left\{ \!
\varphi
\!+\!
\tan^{-1} \!
\left(\!
\tan \! {\displaystyle \frac{f}{2}}
\cos \theta \!
\right) \!
\right\}  .
\label{gaugepotA4}
\eeq
Inversely, from
(\ref{gaugepotA4}),
we can easily derive
(\ref{gaugepotA3})
by noticing the differentiation
given below
\beq
d \!
\left\{ \!
\varphi
\!+\!
\tan^{-1} \!
\left(\!
\tan \! {\displaystyle \frac{f}{2}}
\cos \theta \!
\right) \!
\right\}
\!=
d \varphi
+\!
{\displaystyle
\frac{1}
{1
\!+\!
\tan^2 \! {\displaystyle \frac{f}{2}}
\cos^2 \theta}
} \!\!
\left\{ \!
\sec^2 \! {\displaystyle \frac{f}{2}} \!
\left( \! {\displaystyle \frac{f}{2}} \! \right)^{\! \prime} \!
\cos \theta dr
\!-\!
\tan \! {\displaystyle \frac{f}{2}}
\sin \theta d \theta \!
\right\} \! .
\label{gaugepotdiff}
\eeq


\end{document}